\pdfoutput=1
\documentclass[5p,twocolumn,preprint,a4paper,times,12pt]{elsarticle}

\usepackage{graphicx}
\usepackage{amssymb}
\usepackage{wasysym}   % diameter&arrow symbols.
\usepackage{xfrac}

\usepackage{hyperref}

\journal{}

\begin{document}

\begin{frontmatter}
\title{Measurements of the low energy neutron and gamma ray accompaniment of extensive air showers in~the~knee region of~primary cosmic ray spectrum}

\author[lpi]{A.\,Shepetov}
 \ead{ashep@www.tien-shan.org}
\author[lpi]{A.\,Chubenko}
\author[ipt]{B.\,Iskhakov}
\author[ion]{O.\,~Kryakunova}
\author[iet]{O.\,Kalikulov}
\author[lpi]{S.\,Mamina}
\author[iet]{K.\,Mukashev}
\author[lpi]{V.\,Piscal}
\author[lpi]{V.\,Ryabov}
\author[iet]{N.\,Saduev}
\author[ipt]{T.\,Sadykov}
\author[ion]{N.\,Salikhov}
\author[ipt]{E.\,Tautaev}
\author[lpi]{L.\,Vil'danova}
\author[lpi]{V.\,Zhukov}

\address[lpi]{P.\,N.\,Lebedev Physical Institute of the Russian Academy of Sciences (LPI), Leninsky pr., 53, Moscow, Russia, 119991}
\address[ipt]{Satbayev University, Institute of Physics and Technology,  Ibragimova str. 11, Almaty, Kazakhstan, 050032}
\address[ion]{Institute of Ionosphere, Kamenskoye plato, Almaty, Republic of Kazakhstan, 050020}
\address[iet]{Al-Farabi Kazakh National University, Institute of Experimental and Theoretical Physics, al-Farabi pr., 71, Almaty, Kazakhstan, 050040}
\date{}

\begin{abstract}
Purposeful investigation of radiation fluxes strongly delayed in relation to the main particles front of extensive air shower (EAS) was undertaken at the Tien Shan Mountain Cosmic Ray Station. It was found that the passage of the EAS can be accompanied by the delayed thermal neutrons and by the soft (30--50)\,keV gamma rays, mostly concentrated within a region of about (5--10)\,m around shower axis, where the integral radiation fluence can vary in the limits of $(10^{-4}-1)$\,cm$^{-2}$ for neutrons, and of $(0.1-1000)$\,cm$^{-2}$ for gamma rays. The dependence of signal multiplicity on the shower size $N_e$ has a power shape both for the neutron and gamma ray components, with a sharp increase of its power index around the value of $N_e\approx 10^6$, which corresponds to the position of the $3\cdot10^{15}$\,eV knee in the primary cosmic ray spectrum. Total duration of detectable radiation signal after the EAS passage can be of some tens of milliseconds in the case of neutron component, and up to a few whole seconds for gamma rays. The delayed  accompaniment of low-energy radiation particles can be an effective probe to study the interaction of the hadronic component of EAS.
\end{abstract}


\end{frontmatter}

\section{Introduction}
The study of the neutrons which originate from the interaction of high energy cosmic ray particles with the matter has a rather long history; actually, it can be traced back to 1940s when the paper \cite{bethe1940} appeared in which the flux of environmental neutrons was proposed as a proper messenger on the properties of nuclear interaction in the energy range of a typical cosmic ray particle. At that time this suggestion gained continuation in the early works \cite{tongiorgi1948,tongiorgi1948b,tongiorgi1948c,tongiorgi1948d} which considered particularly neutron signals obtained by registration of extensive air showers (EAS). Since the invention of the world wide neutron monitor network for the global registration of the cosmic ray intensity  variations  in the middle 1960s \cite{simpson,hatton_supermonitor_inbook} the idea to apply this kind of neutron detectors to systematic study of the EAS hadronic component was realized in a number of experimental works which used some monitors as part of shower installation, both at sea level \cite{boehm1969,kozlov1981} and at a high altitude sites \cite{danilova1964,stenkin1999(monitor-mexico),stenkin_icrc2001(monitor-baksan)}. Later on, a neutron monitor based technique was applied systematically in 1990s at the Tien Shan Mountain Cosmic Ray Station, in the frame of the complex \mbox{{\it Hadron}} experiment which was aimed at the investigation of the hadronic component in the EAS with primary energies of $(10^{14}-10^{17})$\,eV   \cite{shalour1993,shalour1999,jopg2001,shalour2001,shalour2003,shalour2005ae}.

Since the initial purpose of a classic neutron monitor was precise measurement of the intensity of energetic cosmic rays, starting from a few GeV order energy and higher, its construction was designed in such a way as to exclude any influence on the part of the local neutron background on the registered counting rate. Correspondingly, any traditionally used neutron monitor set-up includes an effective outer shielding (typically, a layer of light, hydrogen-rich material) which acts as a reflector and absorber of the low energy neutrons from external nearby environment \cite{carmichel_supermonitor}. As a consequence, from the viewpoint of the hadron detection problem the neutron monitor is a rather high threshold detector, inconvenient for investigation of low-energy ($E_h\lesssim$1~GeV) cosmic ray hadrons, and in particular of the low energy neutron flux, in spite of the fact that slow neutrons with kinetic energy of a thermal magnitude order ($\sim$$10^{-2}$\,eV) are present in abundance among the particles connected with the EAS passage. (As, for example, it was shown in \cite{yanke2011}, such neutron monitor like set-ups have only residual ($\leqslant$1\%) probability of neutron detection in the thermal energy range).

On the other hand, since the times when some evidences of an unexpectedly high neutron production within the core region of large sized extensive air showers were obtained in the experiments with the neutron monitor at the Tien Shan Cosmic Ray Station \cite{jopg2001}, a question arose on the behavior of low-energy (thermal and epithermal) neutron fluxes around the EAS center. The urgency of the analysis of low-energy neutron component both in the central region and at the periphery of powerful EAS was discussed in various publications  as well \cite{linsley-subluminal1984,erlyneutrothunder,yakutsk-delayed_pulses2013}. Consequently, a special experiment on registration of the flux of low-energy neutrons was fulfilled at the Tien Shan station in the end of 1990s \cite{jopg2008}. Together with the detection of the neutron component, the configuration of that experimental installation included a nu\-m\-ber of gamma ray detectors. It was found that neutron signal from the passage of a large size EAS cores was often accompanied by an additional flux of gamma radiation, the origin of which was at that time ascribed to a capture process of thermalized neutrons by atomic nuclei in the local environment. It should be noted that at those days there was not any operating shower installation at the Tien Shan station, so that both the axis position and the size of EAS were roughly estimated by the spatial distribution and registered density of the neutron and gamma ray particles.

The application of the thermal neutron registration technique to the studies connected with the physics of extensive air sho\-wer was considered in detail in the publications \cite{stenkin2002,stenkin-no-knee,stenkin2007} were a specialized scintillation neutron detector was proposed for the purpose. This detector is based on a thin layer of the \mbox{ZnS(Ag)} and \mbox{$^6$LiF} (or \mbox{$^{10}$B$_2$O$_3$}) alloy grains covered with a thin transparent plastic film. Such kind of neutron detectors is very convenient for mass application in the large scale shower installations because of its cheapness, production simplicity, and possibility of simultaneous registration both of the neutron and electron components of EAS in a single detector, with their further effective separation by the amplitude and time profiles of scintillation flash. Up to the present time, a prototype of suggested detector has been tested in a number of EAS experiments, and spatial and temporal characteristics of the thermal neutron flux which accompany the EAS passage were obtained both underground \cite{stenkin-bacsan2009}, at sea level \cite{stenkin-mephi2011,stenkin-mephi2013,stenkin2013,stenkin-mephi2016}, and in the mountains \cite{stenkin-yangbajing2016}. Currently, a systematic use of such neutron detectors is anticipated at the modern large size EAS array LHAASO \cite{stenkin-lhasoo2019}.

Since the year 2015, the EAS particles detector system of the Tien Shan mountain station was put once again into operation state after its prolonged modernization period, and regular registration of the neutron component of extensive air showers was resumed here at modern level of the experimental technique \cite{our2017_stm32}. The basic goals of the present Tien Shan cosmic ray experiment were listed in the program article \cite{ontien-nim2016}; in particular, it was stated there that application of the various kinds of neutron detectors together with their specific data collection procedure gives a unique possibility to register the response of a single experimental installation to hadronic interactions in an exceptionally wide range of energy deposits, starting from the thermal neutrons, and up to a few TeV order energies of the cosmic ray hadrons. This is especially important for the detection of intensive radiation flows in the central region of powerful air showers. In turn, precise investigation of the core EAS region with concentration of its most energetic particles is a key condition for collection of the new experimental data which can finally help to solve the long standing problem of the knee in the energy spectrum of the primary cosmic rays, that is of the sharp change of its power index around the energy of $3\cdot 10^{15}$\,eV.

Besides the neutron signal, it was realized now at the Tien Shan station a continuous systematic detection of gamma rays with a number of energy thresholds, the lowest of which corresponds to the soft radiation limit of $E_\gamma\approx 30$\,keV. Primary reason to include the gamma detectors into Tien~Shan experimental complex was the consideration that if soft radiation does actually result from the capture of thermalized evaporation neutrons which originate in the interaction of the nuclear active cosmic ray components, this radiation can be another messenger on the properties of high energy nuclear reactions, being more abundant than the neutron component itself. In contrast to the situation of the former work \cite{jopg2008}, presently both the neutron and gamma detectors at the Tien Shan station operate in strict synchronization with the shower detector system which permits to define basic EAS characteristics (above all, the shower size $N_e$ and the precise position of shower axis) together with the intensity of accompanying neutron and gamma ray signals in every detected EAS event.

The aim of the current publication is to present new  experimental data on the properties of low-energy neutron and gamma ray fluxes which accompany the EAS passages in the $(10^{14}-10^{17})$\,eV range of the primary cosmic ray spectrum, and on their dependency on the characteristics of corresponding shower. The measurements were made at the height of 3340\,m above the sea level by the particles detectors complex of the Tien Shan Mountain Cosmic Ray Station.

The structure of this paper the following. In the next Section~\ref{sectiista} it is given the review  of the basic technical facilities which were applied in the discussed experiment: the Tien Shan system of particles detectors for the registration of the charged component of extensive air showers, and the detectors of low energy neutrons and soft gamma radiation. Then, Section~3 follows which is aimed at presentation of the newly obtained experimental data on the delayed neutron accompaniment of extensive air showers. In this section, firstly, it is considered the temporal distribution of the neutron signals which were detected after the passage of the EAS front, and, secondly, analytical approximation of this distribution is used to calculate the total integral fluence of the accompanying neutron flux. Similarly, Section~4 deals with temporal and multiplicity characteristics of the delayed EAS connected gamma radiation, as well as with particularities of its energy spectrum. Both in the third and fourth sections a special attention is drawn to the fact that some drastic change does exist in dependence of the average multiplicity of delayed neutron and gamma ray signals on the size $N_e$ of corresponding air shower, which resides around the value of $N_e\approx 10^6$. It is noted there that with application of a standard $N_e\rightarrow E_0$ recalculation rule which was used in previous shower experiments at the Tien Shan station for transition from an EAS size to its primary energy $E_0$  \cite{ontien_icrc1987__e0_through_ne_ru,hadron_spc_2017}, the border value $N_e\approx 10^6$ corresponds to the $E_0\approx 3\cdot 10^{15}$\,eV point on the energy scale, that is just to position of the prominent knee in the primary cosmic ray spectrum. Experimental facts presented in the whole article are to be summed up in Conclusion section where it is pointed out that the existence of a rather intensive ``afterglow'' of delayed low energy neutrons and gamma rays was now revealed within the core region of powerful EASs. Since both these components must evidently originate from the interaction of energetic EAS hadrons with environmental matter, this effect can be a substantial source of information on the properties of high energy hadronic interactions which still remain, to a considerable extent, unclear.

\section{Instrumentation}
\label{sectiista}

\subsection{The EAS registration technique}

\begin{figure*}
\begin{center}
\includegraphics[width=0.8\textwidth, clip, trim=0mm 11mm 0mm 11mm]{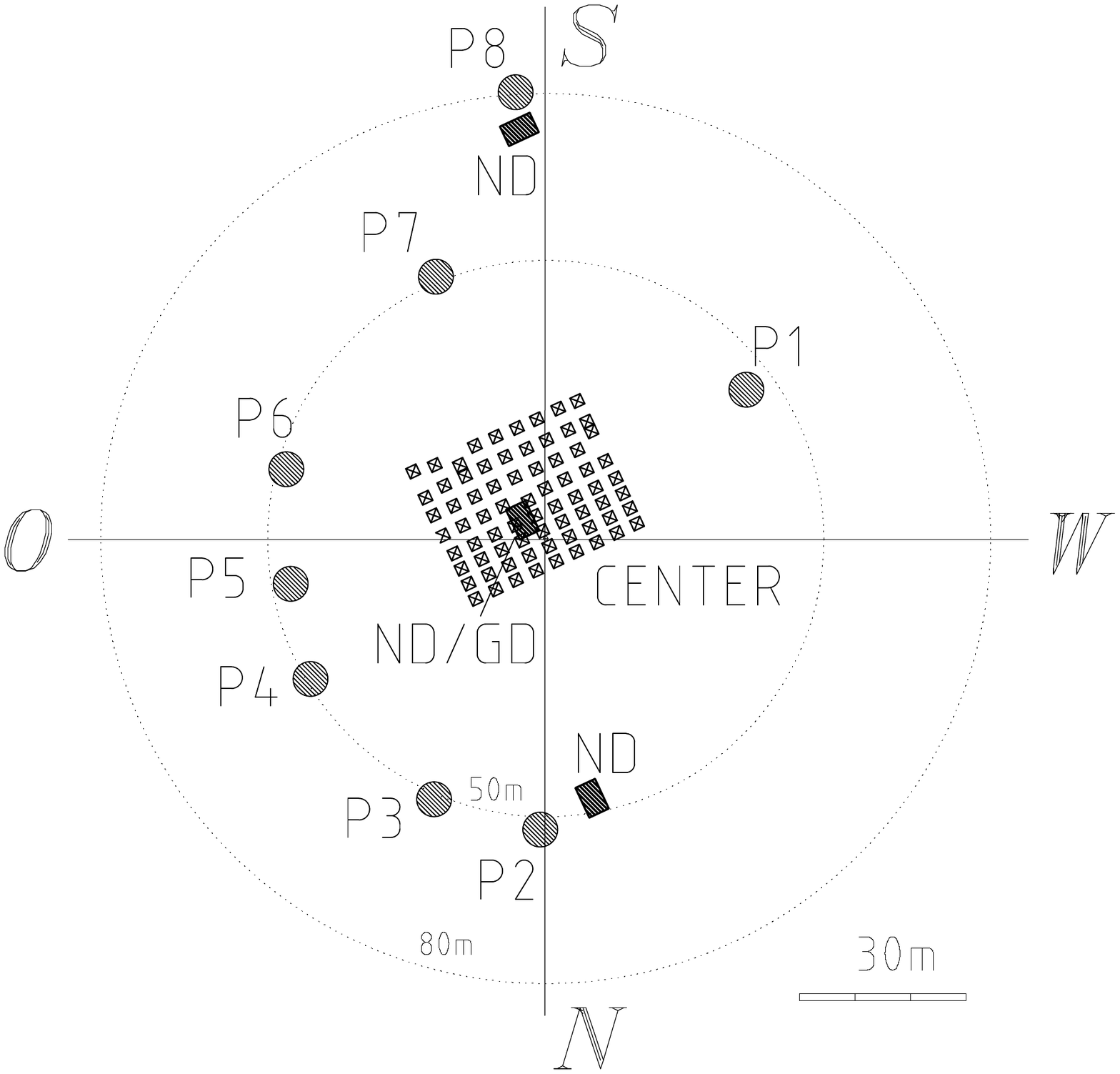}
\caption{Configuration of the central part of Tien Shan shower detector system at the time of considered experiment. \textit{CENTER}---the central ``carpet'' of scintillation charged particles detectors (which are shown with small separate squares); \textit{ND/GD}---the detectors of the low-energy neutrons and gamma rays; \textit{P1}--\textit{P8}---the peripheral EAS particles detectors.}
\label{figiscintidete}
\end{center}
\end{figure*}

The Tien Shan shower installation is designed for EAS detection in the range of the primary cosmic ray energies of $E_0\approx (10^{14}-10^{17})$\,eV.
As it is shown in Figure~\ref{figiscintidete}, the central part of this installation is an array of 72~scintillation charged particles detectors spread over a rectangular 30$\times$30\,m$^2$ ``carpet'' area, with the steps  between neighbouring points being of 3\,m and 4\,m along two orthogonal directions. Such a rather dense detector disposition in the region close to the installation center was selected to investigate as precisely as possible spatial distribution of the flow of charged particles  within the EAS cores \cite{ontien-nim2016}.  At the time of the considered experiment 8~additional detectors (P1-P8 in Figure~\ref{figiscintidete}) were spread nearly circumferentially around this ``carpet'' at distances of about $50-70$\,m from its center for the detection of particles at the EAS periphery.

The main sensitive part of the EAS particles detector used at the Tien Shan station is a $0.5\times 0.5\times 0.05$\,m$^3$ poly\-sty\-re\-ne scintillator block placed at the bottom of a $\sim$0.5\,m high py\-ra\-mi\-dal light-tight\ \ reflector, with a photomultiplier tube (PMT) installed at its top. In the case of EAS passage the amount of relativistic charged particles which have come through scintillator block, the peak intensity of a scintillation light flash, and the amplitude of an electric pulse on PMT output occur being proportional to each other (assuming the linear amplification mode of PMT operation). Hence, a digitization of the PMT output signal permits to define the local density of the EAS particles in the point of detector disposition. This latter task is solved by a multichannel amplitude to digital converter system which was especially designed for the modern Tien Shan cosmic ray experiment. At present time the electronic channels of this system ensure a saturation free measurement of the density of EAS particles in the limits of $(1-10^5)$\,particles/m$^{2}$ in each detector point \cite{ontien-nim2016}, which permits to study precisely the particles density distribution around the core region of up to $E_0\approx 10^{16}$\,eV EAS.

In turn, spatial distribution of the particles density over the area of the whole detector system can be used for the estimation of the main EAS characteristics, primarily of the position of shower axis in the plane of installation, and of the total  charged particles number in the shower---the EAS ``size'' $N_e$. Detailed discussion of the  algorithms of particles density calculation by the values of scintillation amplitudes, and of restoration of the EAS parameters by these data is given in \cite{ontien-nim2016}. Practically, at the time of the considered experiment the accuracy of the EAS axis position determination for the showers which fall within the limits of the central detector ``carpet'' was below the spatial step between its neighbouring points, \textit{i.\,e.} it was of at least 3--4\,m or better. The relative error of the shower size determination in these events was about 20--30\%.

Apart from the measurement of the density of EAS particles, the system of shower detectors is used for elaboration of a control trigger pulse signal which marks the arrival time of an EAS. Further on, this trigger is transmitted to all detector subsystems of the Tien Shan station for synchronization of their data sampling processes. For generation of shower trigger the momentary output signals of all 72~detectors which form the central ``carpet'' part  are summed together continuously during the whole operation time of shower installation. This summation is made in analog form by a mul\-ti\-chan\-nel circuitry built of an array of operational amplifiers, and at the moment when the level of this sum signal exceeds some pre-defined threshold an analog amplitude discriminator generates the trigger pulse. In such scheme the typical resolution time of the trigger system is defined mostly by duration of the shaped output pulses of individual particles detectors. In considered experiment this latter value was kept at about 2--3\,$\mu$s in all detectors.

The lowest shower size limit $N_e^{min}$ of the effective EAS selection with the described trigger system can be simply regulated by tuning the threshold of the trigger discriminator; at the time of the considered experiment it was set in such a way that the showers with $N_e\gtrsim 3\cdot 10^5$ (\textit{i.\,e.}\,\,with primary energy of $E_0\gtrsim 10^{15}$\,eV) whose axes were hitting the central detector ``carpet'' were registered with a nearly 100\%~probability, but the detection of smaller showers up to $N_e\gtrsim (3-5)\cdot 10^{14}$\,eV was possible as well with a somewhat reduced efficiency.
% integral intensity at N_e>3\cdot 10^5 is 10^{-2}/m^2/h
% >>> 1.e-2*(30*30)*4200
%37800.0
%>>> 850000 / (1.e-2*(30*30)*4200)
% 22.486772486772487 <-- this is the reserve

\subsection{Neutron detector}

\begin{figure*}
\begin{center}
\includegraphics[width=0.49\textwidth, trim=0mm 0mm 0mm 0mm]{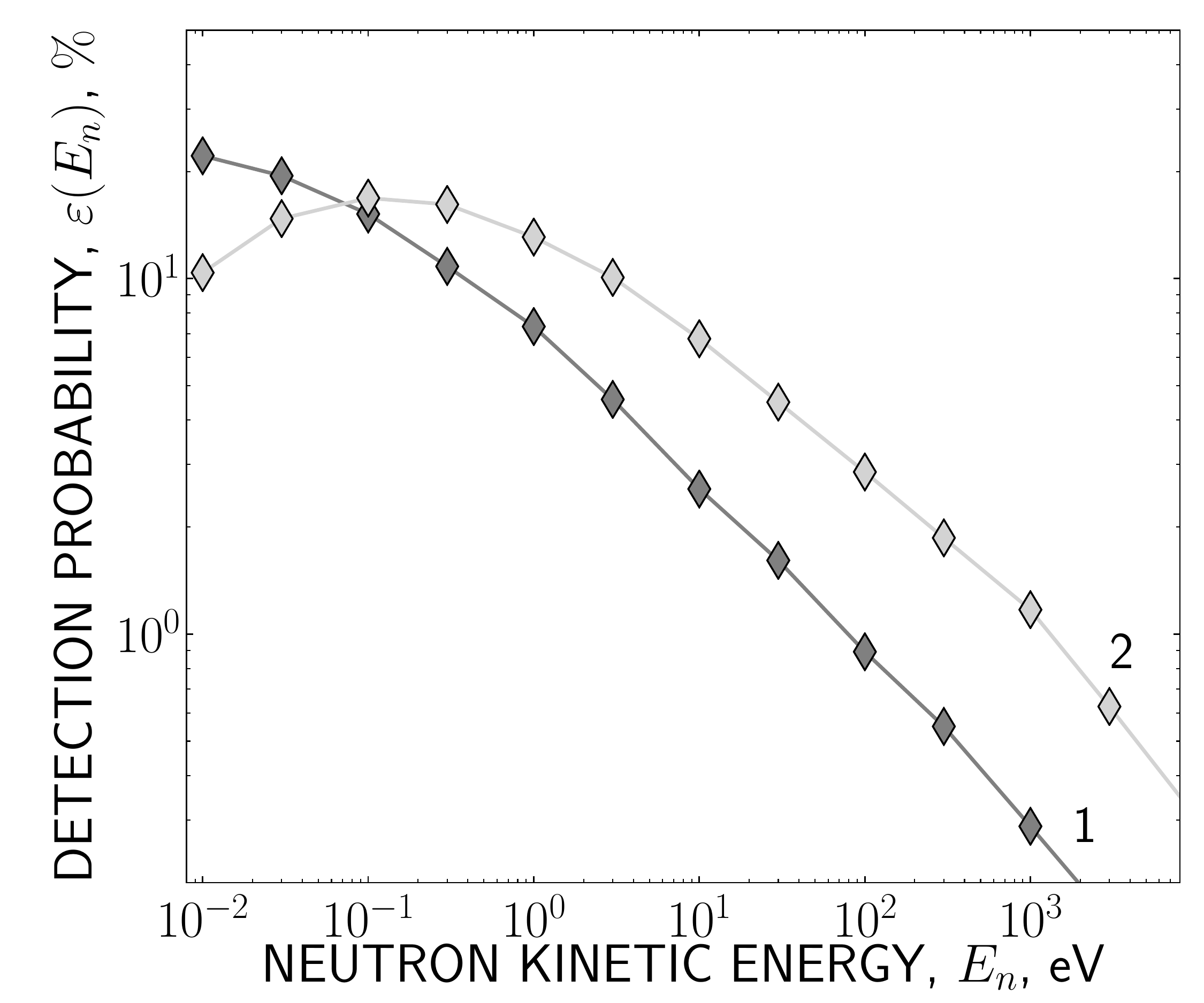}
\includegraphics[width=0.49\textwidth,trim=0mm 0mm 0mm 0mm]{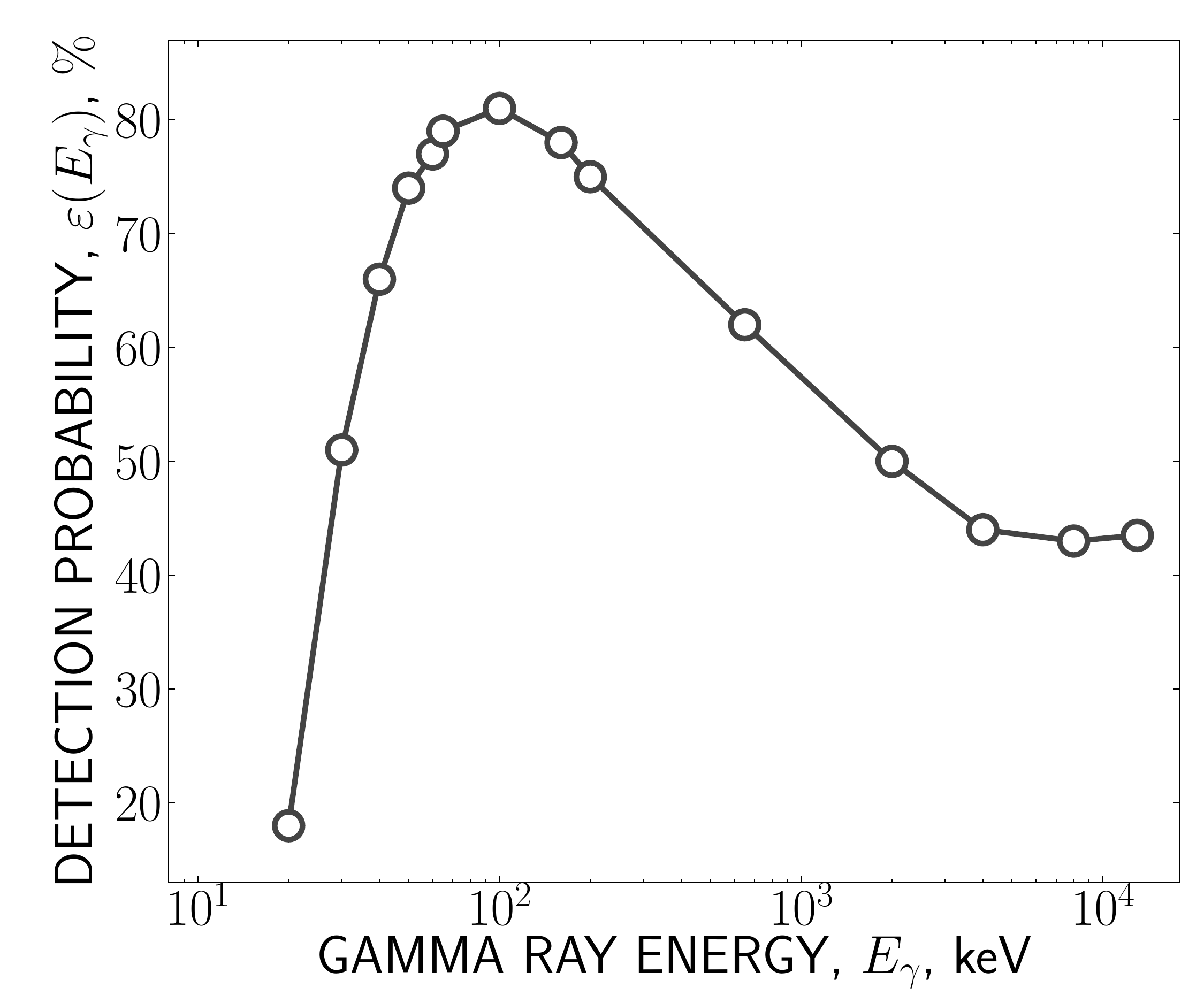}
\end{center}
\caption{\label{figieffici} Energy dependence of the particles detection probability (that is the detector efficiency) in the Tien Shan experiment---the results of the Geant4 simulations made for a bare neutron counter {\it (1)}, for a counter surrounded with a 0.6\,cm thick PVC moderator {\it (2)}, and for a gamma ray detector (second plot).}
\end{figure*}

The detection technique of the low-energy neutrons presently used at the Tien Shan mountain station is based on the $\diameter 30\times 1000$\,mm$^2$ gas discharge counters filled with a mixture of natural argon and enriched $^3$He, so the neutron detection is possible there due to nuclear reaction $n(^3$He,$^3$H$)p$. Since the effective cross-section of this reaction diminishes rapidly with the rise of interaction energy some counters can be surrounded by a hydrogen rich moderator material (po\-ly\-vi\-nyl\-chlo\-ri\-de (PVC)) to extend their sensitivity into a higher energy range of incident neutrons.

Energy dependence of the probability of nuclear reaction, that is of the efficiency of neutron detection was estimated through a complete Monte Carlo simulation of the neutron interaction processes with internal material of the counter and surrounding moderator. For this purpose on the basis of the Geant4 toolkit \cite{geant_base} it was designed a program model of neutron detector. This model takes into account both the characteristic features of a real neutron counter (its geometry, the presence or absence of moderator, the composition of the gas filling, \textit{etc}), and the typical conditions of the outer environment at the Tien~Shan mountain station (primarily, the presence and atomic content of the soil beneath the detector, then the usual atmospheric pressure and humidity values, \textit{etc}). Complex model of particles interaction used in this simulation includes a set of elementary interaction processes provided by Geant4:
\begin{itemize}
\item[-]
elastic neutron coincidence processes for a wide range of kinetic energy of incident neutrons, starting from the thermal energy order values and up to 20\,MeV;
\item[-]
inelastic interactions of the thermal, epithermal, intermediate, and high energy neutrons;
\item[-]
the process of radiative neutron capture, so as standard electromagnetic interactions of the resulting gamma ray quanta: the $e^\pm$-pair production, Compton and photoelectric effects.
\end{itemize}

By simulations, the model detector was put into an isotropic flux of the neutrons with a fixed kinetic energy $E_n$: before to start the trajectory of a next probe particle its initial position was selected in a random point somewhere on the outer surface of the model detector, then three directional cosines were set also randomly to define the direction of its momentum, then the simulation of the particle's trajectory begun. Since in reality we have to deal mostly with the flux of thermalized neutrons which diffuse evidently uniformly in the outer environment around the detector, such geometry of simulation seems to be quite adequate to the configuration of the real experiment.

By simulation, the fact of neutron detection was signaled by an appearance of $^3$H/$p$ particles pair inside the internal volume of model counter. In turn, overall registration probability of the neutrons with kinetic energy $E_n$ was defined as a relation of the number of ``detected'' neutrons to the total amount of the primaries with given $E_n$ which were included into simulation. The results of this calculation are presented in the first plot of Figure~\ref{figieffici}.

The registration procedure of neutron flux intensity in the discussed experiment consists of counting of the electric pulse signals which come from each neutron counter separately during a fixed time period (a gate time) after the passage of the extensive air shower. After proper shaping, the pulse signals from the anode wires of all the counters are transmitted to the microcontroller registration unit \cite{our2017_stm32} which can calculate the amount of input pulses obtained in a set of succeeding 250\,$\mu$s long time intervals. The starting point of the whole data recording sequence is synchronized with an EAS trigger signal provided by the central shower installation. The microcontroller driver program runs continuously in a pre-trig\-ger/post-trig\-ger operation mode, so the records of neutron signal intensity are available with a 250\,$\mu$s resolution both before and after the EAS passage. The number of succeeding time intervals in the ``post trigger'' part of the sequence is 4000, so the sum duration of a signal intensity record in each EAS event amo\-unts to one second. In parallel with such a time series operation mode which is strictly bound to an EAS trigger, the microcontroller driver program ensures regular monitoring of the background intensity for all connected signals which goes on continuously and fully independently of any external events. A more detailed description of the neutron signal registration procedure accepted in the Tien Shan cosmic ray experiments is given in publications \cite{our2017_stm32,ontien-nim2016}.

At the time of the considered experiment three detector sets each including twelve neutron counters were used for the detection of the neutron accompaniment of EAS. As it is shown in Figure~\ref{figiscintidete}, one of these detectors (\textit{ND/GD}) was placed near the center of the shower detector system, and two others (\textit{ND})---at distances of about 50\,m and 80\,m from this point. Such disposition of the neutron detector sites allowed to measure the EAS connected neutron flux simultaneously both around the core and at periphery of each detected shower.

\subsection{The gamma radiation detector }

On the basis of preliminary experiment \cite{jopg2008}, some surplus of soft (tens of keV---3~MeV) gamma radiation was apriori expected to exist above its usual background after EAS passages, especially around the region of shower core. This was the reason to install a special low-thre\-shold gamma ray detector together with the neutron one, in the same point \textit{ND/GD} near the center of the Tien Shan shower system. The gamma detector used for this purpose consists of NaI(Tl) crystal scintillator coupled with photomultiplier tube (PMT). The scintillator is of \diameter 110$\times$110\,mm$^2$ cylindrical shape; together with PMT it is placed inside an aluminum casing with a 1\,mm wall thickness.

To define detection efficiency in various energy ranges of incident radiation it was made a Geant4 simulation of gamma rays propagation within a scintillator crystal. By calculations, initial gamma ray quanta with a fixed energy $E_\gamma$ and isotropic angular distribution (which was organized in the same way as in the above case of the neutron detector simulation) were hitting a model scintillator. The properties of this model---its atomic contents, density, geometrical shape, \textit{etc}---were the same as by real NaI crystal. The model of physical interactions used in  simulation included $e^\pm$-pair production process and processes of Compton and photoelectric effects. The final goal of simulations made for each energy $E_\gamma$ of the incident radiation was to define the corresponding relative share $\varepsilon$ of simulated events in which an absorption of initial gamma quantum occurred within the inner volume of the model scintillator, with simultaneous appearance there of charged product particles capable to cause a flash of scintillation light. It is this parameter $\varepsilon$ which was accepted in the current experiment as the probability of radiation detection, or in other words as the efficiency of gamma detector. Hence, neither the probability of scintillation generation by an excited atom in the crystal nor the quantum efficiency of the PMT photocathode were taken into consideration in these simulations, so the said $\varepsilon$ values should be considered as a possible upper limit of detection efficiency. The resulting dependence  $\varepsilon(E_\gamma)$ is plotted in the second frame of Figure~\ref{figieffici}.

By registration of gamma ray accompaniment of extensive air showers in Tien Shan experiment the detector was operating in the pulse counting mode. Output electric pulses from the PMT anode, with an amplitude being proportional to the energy of every particular gamma ray quantum absorbed in scintillator were transmitted to a set of pulse discriminators with the stepwise increasing operation thresholds. At the time of the considered experiment, twelve amplitude discriminators were used for the purpose with their thresholds set equivalently to consecutive radiation energy increase in the limits from 30\,keV and up to 2000\,keV. Standard digital pulses from discriminator outputs were connected to the same microcontroller driven data registration board together with neutron signals, where they were operated by the same driver program and in the same manner, \textit{i.\,e.} providing two parallel datasets: both the 250\,$\mu$s resolution time series of the input signal intensity with external synchronization by an EAS trigger, and continuous background counting rate measurements in regular monitoring mode.

\section{The neutron accompaniment of extensive air showers}

\subsection{Temporal behaviour of the neutron flux}
\label{sectineutrotempo}

\begin{figure*}
\begin{center}
\includegraphics[width=0.43\textwidth, clip, trim=8mm 11mm 4mm 0mm]{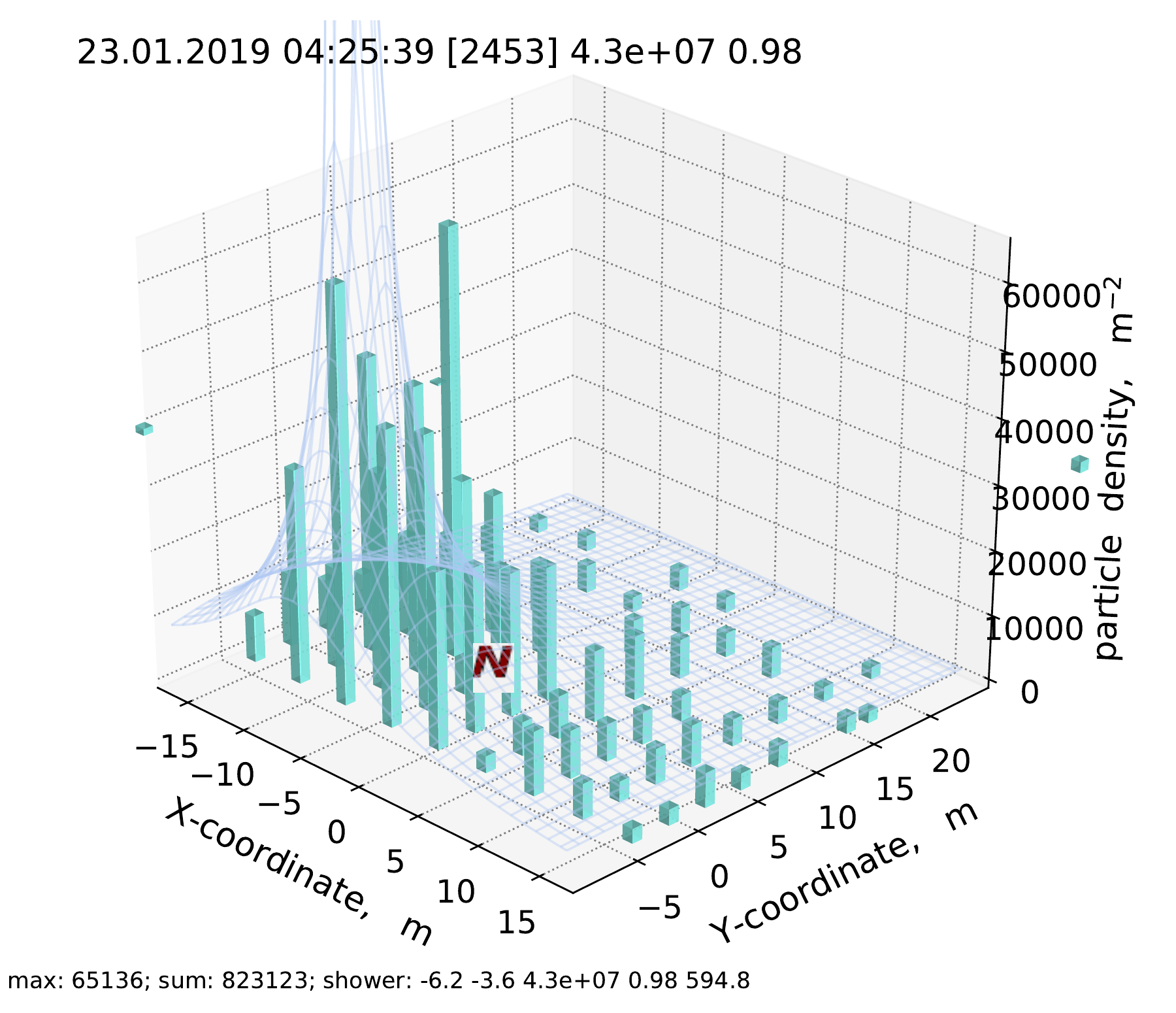}
\includegraphics[width=0.43\textwidth, trim=0mm 0mm 12mm 0mm]{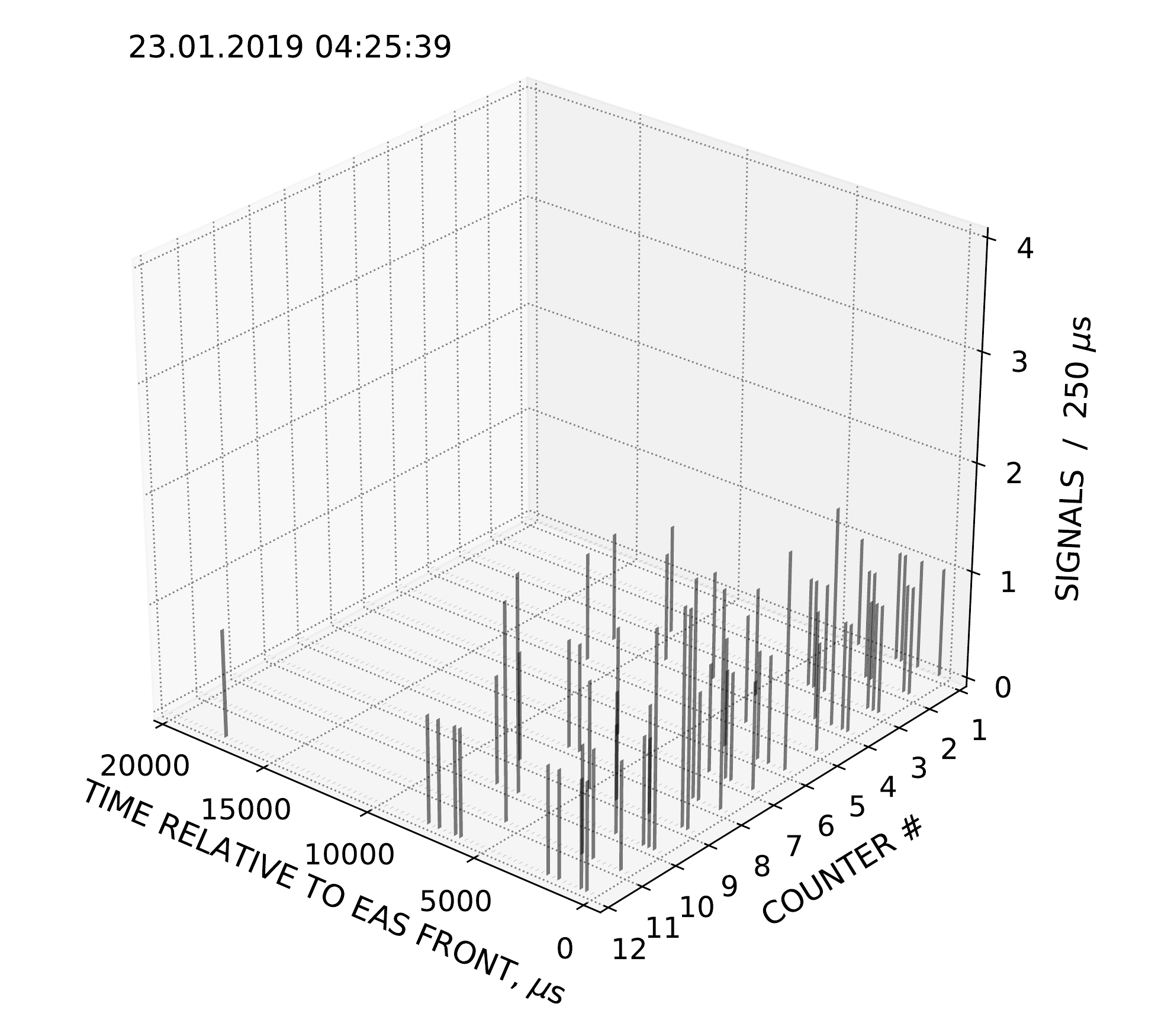}
\\
\includegraphics[width=0.43\textwidth, clip, trim=8mm 11mm 4mm 0mm]{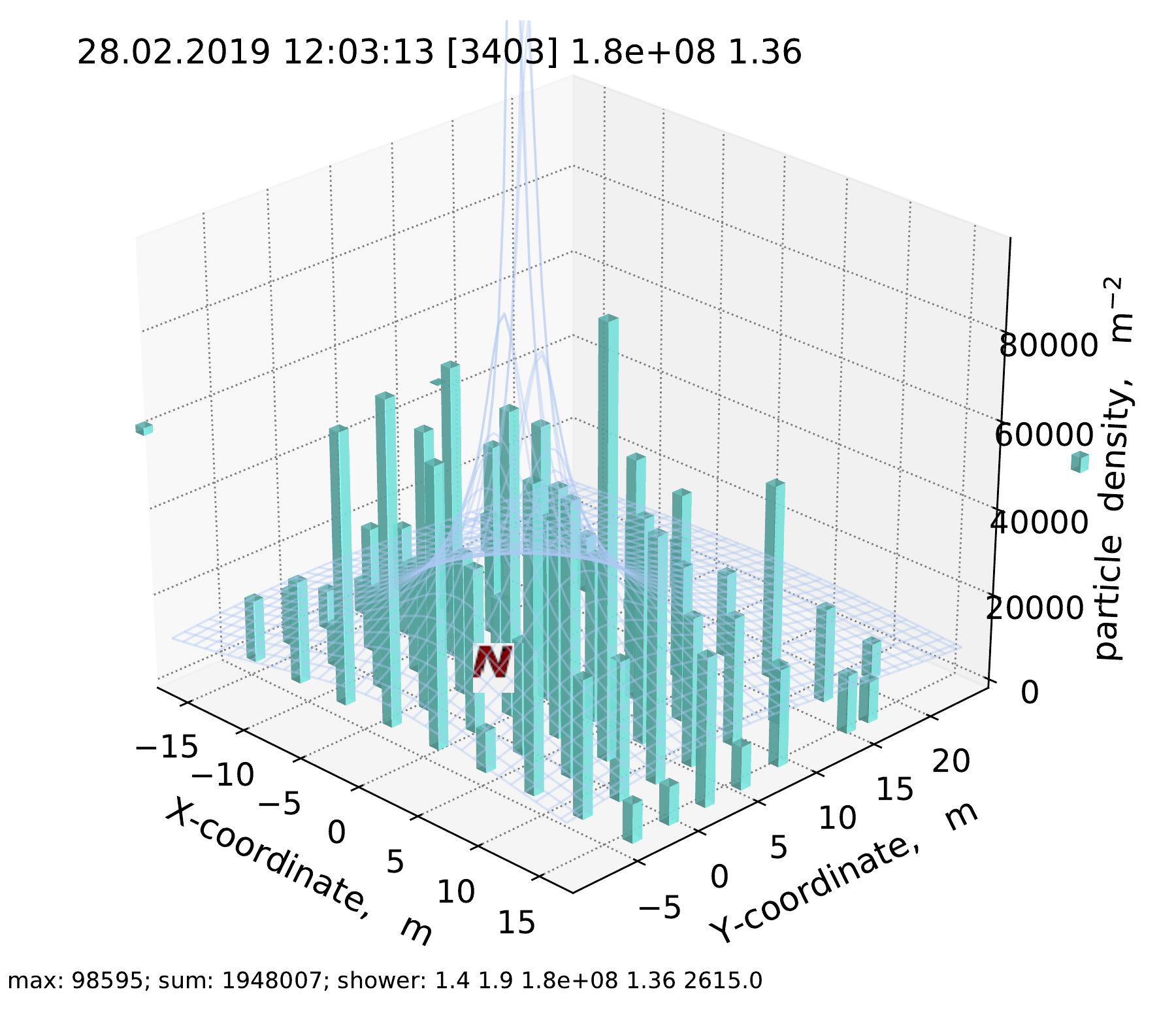}
\includegraphics[width=0.43\textwidth, trim=0mm 0mm 12mm 0mm]{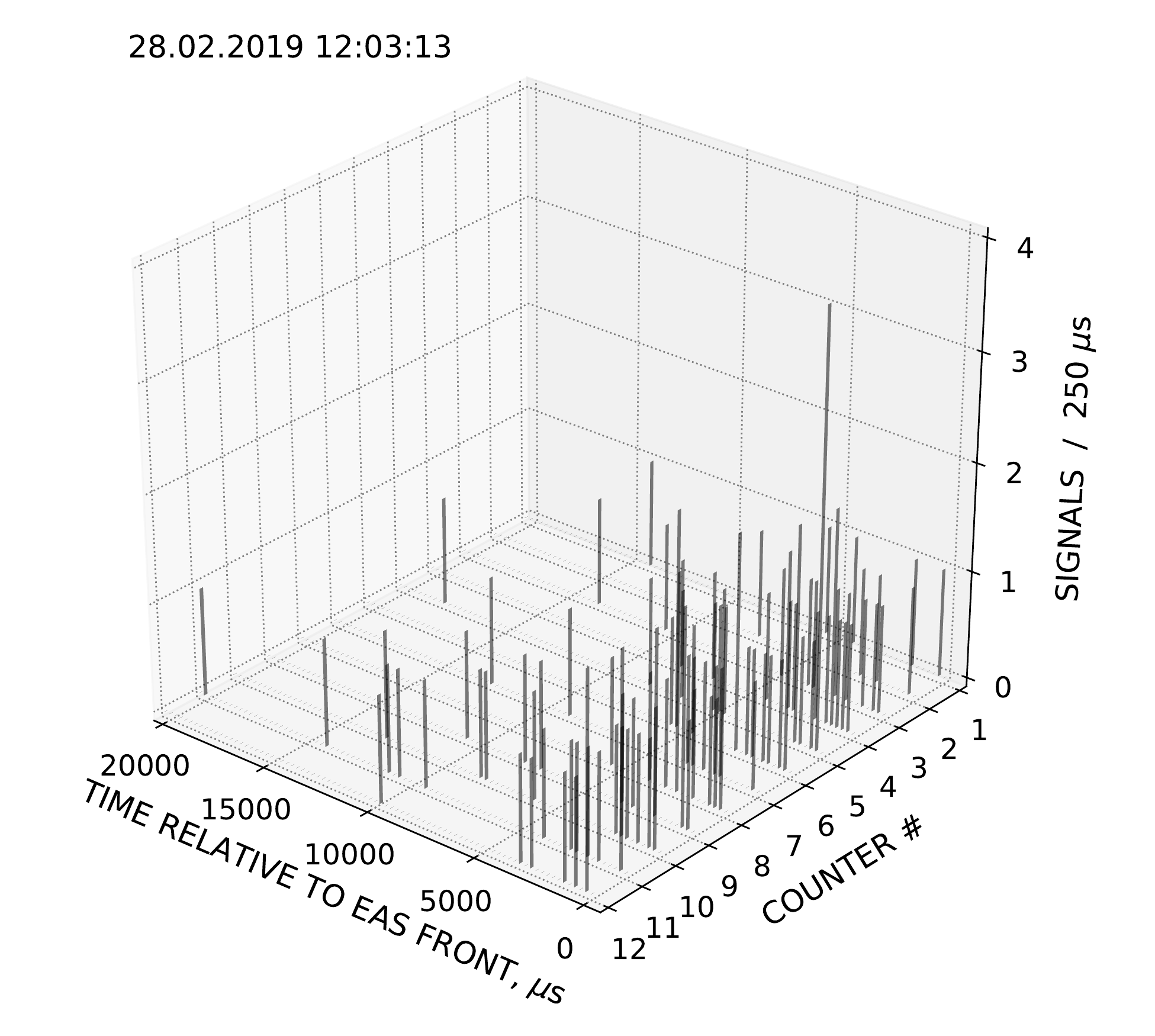}
\\
\includegraphics[width=0.43\textwidth, clip, trim=8mm 11mm 4mm 0mm]{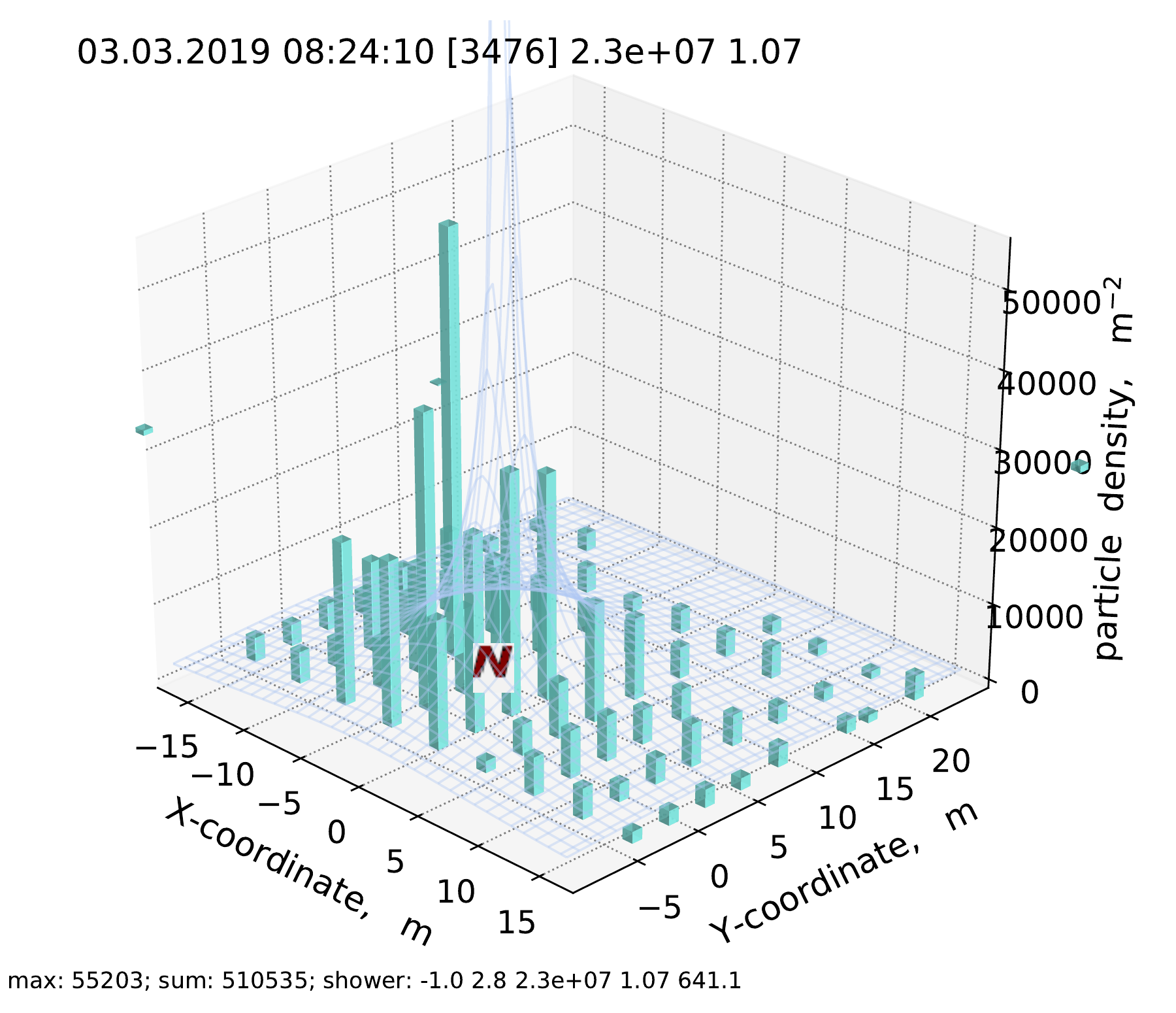}
\includegraphics[width=0.43\textwidth, trim=0mm 0mm 12mm 0mm]{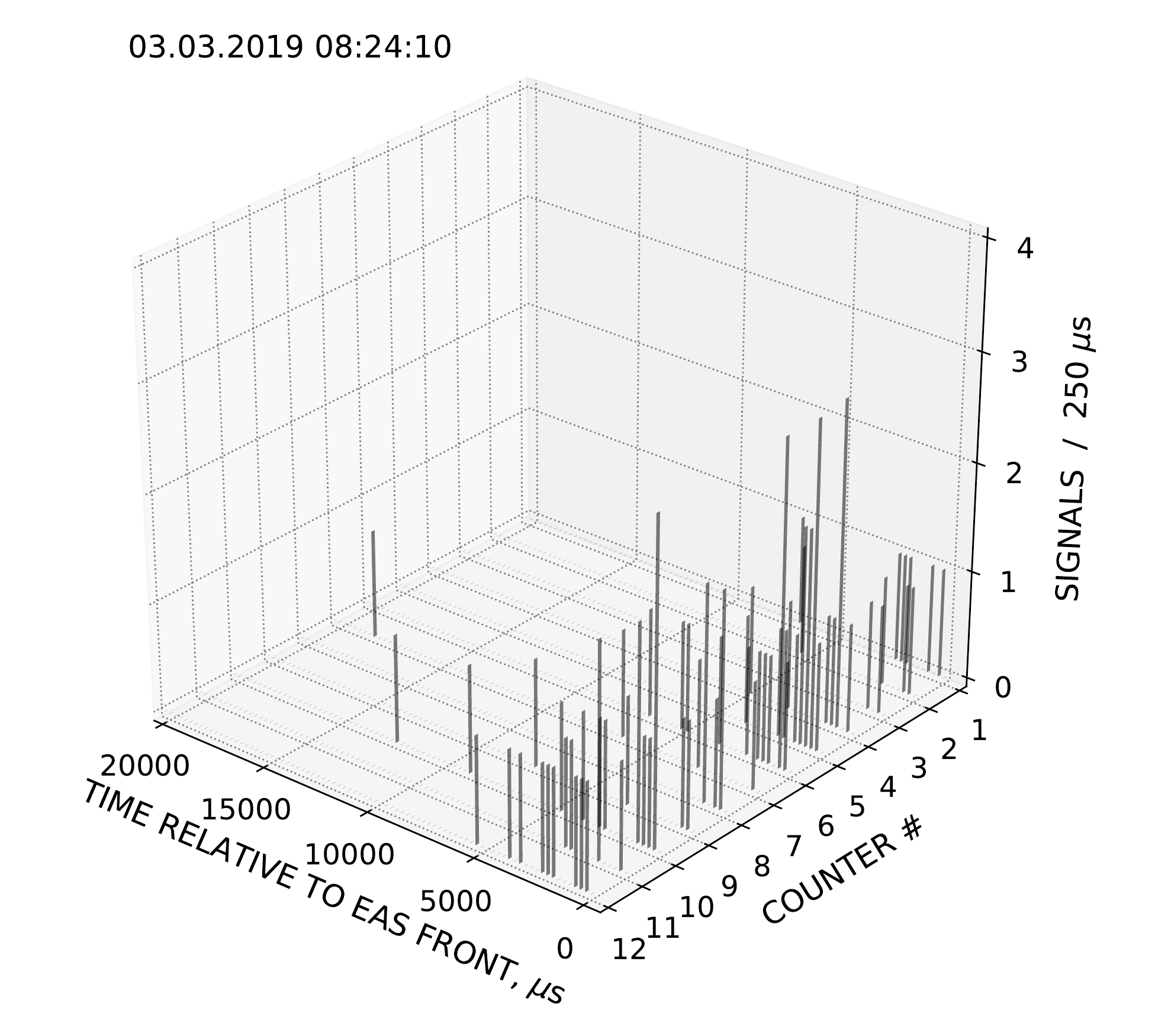}
\caption{Typical EAS events with a delayed neutron accompaniment within the shower core region. Left column: spatial density distribution of the charged shower particles over the plain of the particles detectors system (the columns height is proportional to the measured values of the local particles density, a smooth wireframe surface indicates their 2D approximation, and the $N$ letter marks the position of the central neutron detector). Right column: temporal distribution of the pulse signals from twelve neutron  counters of central detector which have been registered after an EAS passage.}
\label{figineutroevents}
\end{center}
\end{figure*}

The data presented further on were obtained during a nearly 5000\,h long period of simultaneous operation of the Tien Shan shower installation with neutron and gamma ray detectors which was held in the years 2018--2019. In this time interval it was registered about 380k~EAS events with definable parameters. It was found that nearly in half of these cases ($\sim$150k) some signals from neutron counters were detected which were situated just after an EAS passage, but with noticeable delay in relation to the common front of the charged shower particles. In spite of the time gap, these signals were evidently connected with the shower, since their sum intensity remained deliberately above the usu\-al background level during a rather prolonged time, of the order of some tens of milliseconds after the shower.
%
%shescinti=> select sum(timelt)/3600./100 from scintiellitrg where datep>='30.10.2018' and datep<='15.05.2019';
% 4192.6400611111111100
%
%shescinti=> select count(*) from scintiellitrg where datep>='30.10.2018' and datep<='15.05.2019';
% 848585
%
%sheshower=> select count(*) from neutroelli2016b where datep>='30.10.2018' and datep<='15.05.2019' and (mulineu+mulineumode > 0 );
% 347464
%
%
%'with definable parameters'
%
%shescinti=> select count(*) from scintiellishwtrg where datep>='30.10.2018' and datep<='15.05.2019';
% 383891
%select count(*) from ( select neutroelli2016b.datep,neutroelli2016b.timeut,mulineu,mulineumode from neutroelli2016b,ppp where neutroelli2016b.datep=ppp.datep and neutroelli2016b.timeut=ppp.timeut ) as ppp  ;
% 357477
%select count(*) from ( select neutroelli2016b.datep,neutroelli2016b.timeut,mulineu,mulineumode from neutroelli2016b,ppp where neutroelli2016b.datep=ppp.datep and neutroelli2016b.timeut=ppp.timeut and (mulineu+mulineumode > 0) ) as ppp;
% 150466

A sample of neutron events observed in the central detector point $N$ which is placed just in the middle of the shower detector system is shown in the plots of Figure~\ref{figineutroevents}. As it follows from these pictures, an excess of the neutron flux was found in cases when a shower axis was coming within the distance $r\lesssim 5-10$\,m from detector point, and the showers happened to have sufficiently big sizes ($N_e\gtrsim 10^7$, \textit{i.\,e.} $E_0\gtrsim 10^{16}$\,eV). Later on, it was revealed in the Tien Shan experiment that the combination of a large size EAS and close axis location is generally a necessary requirement for observation of such ``neutron accompaniment'' events.

\begin{figure*}
\begin{center}
\includegraphics[width=0.49\textwidth, trim=0mm 0mm 0mm 0mm]{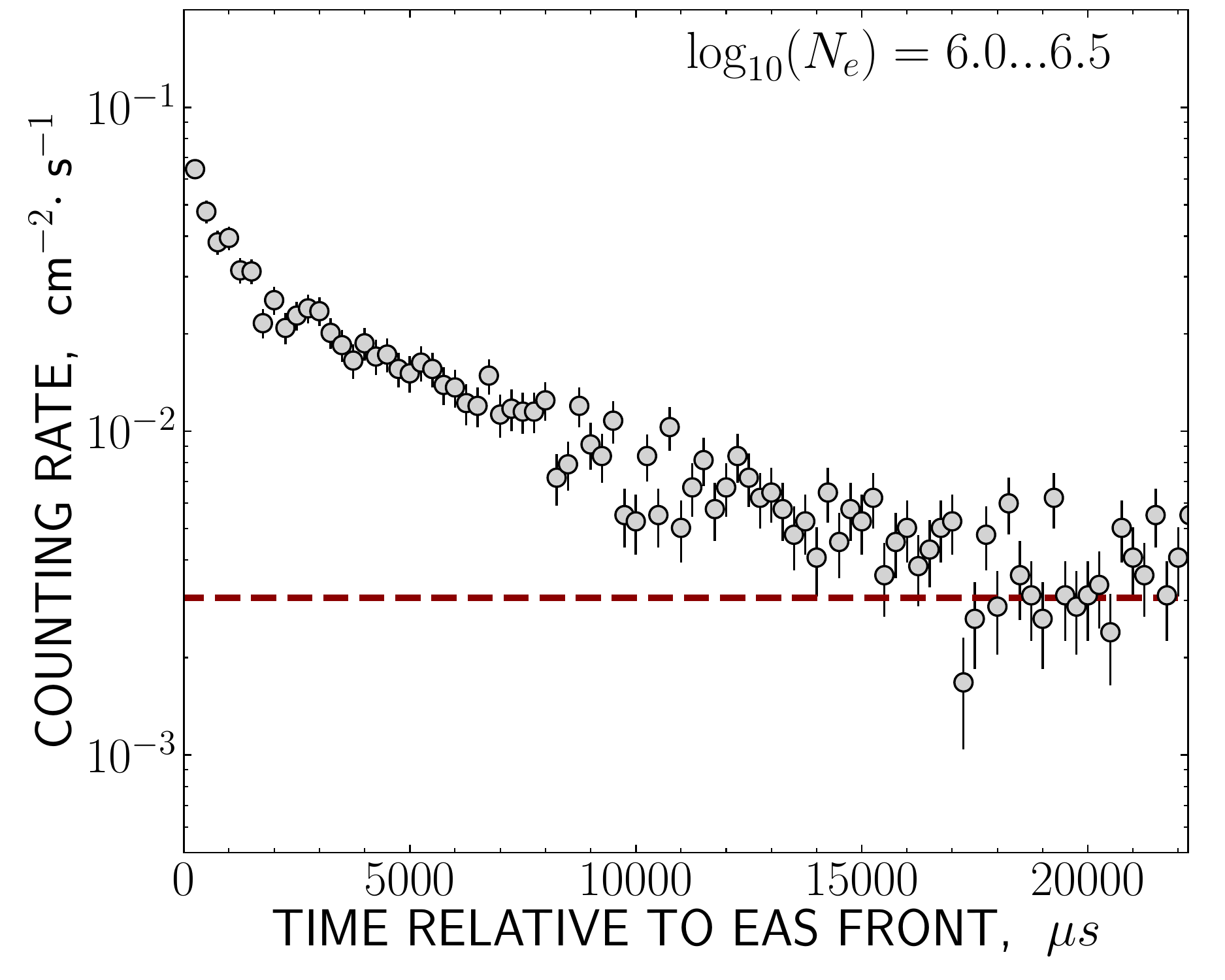}
\includegraphics[width=0.49\textwidth, trim=0mm 0mm 0mm 0mm]{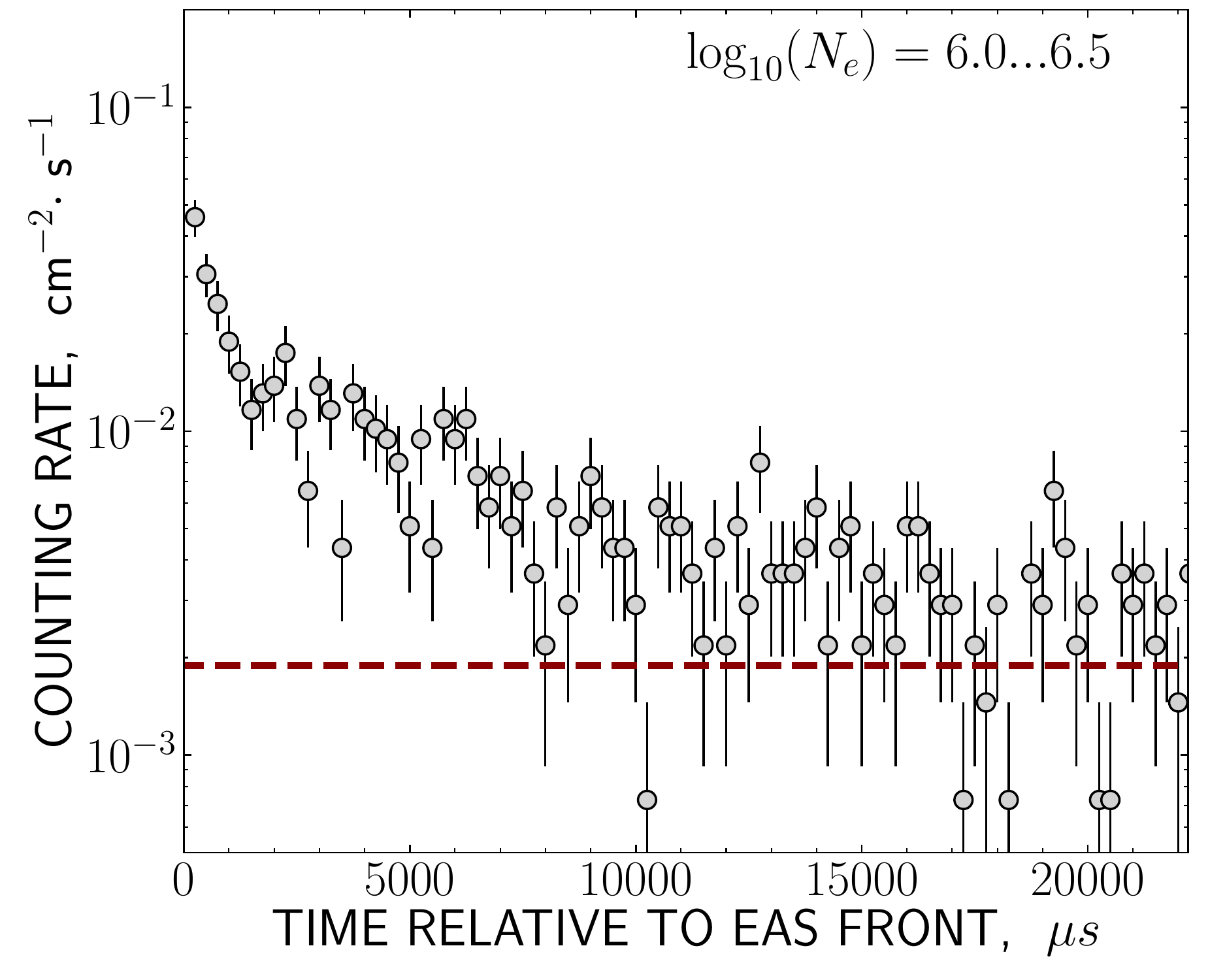}
\caption{Average time distribution of the intensity of low-energy neutron flux registered within the central region ($r\leqslant 10$~m) of the $N_e=10^{6.0}-10^{6.5}$~EAS (with corresponding primary energy $E_0\approx (3\cdot 10^{15}-10^{16})$\,eV). Left plot corresponds to detector build on the bare neutron counters, right plot---to the counters with a 0.6~cm thick PVC moderator. Dashed lines indicate the level of background count intensity for both detector types, zero points on horizontal axes correspond to the time of shower passage.}
\label{figineutrotempo}
\end{center}
\end{figure*}

More systematic and statistically abundant stu\-dy of the delayed neutron accompaniment in EAS can be made by considering the mean characteristics of the detected neutron flux, averaged over a group of shower events with similar values of their $r$ and $N_e$ parameters. As a starting point for such investigation one can use temporal distribution of the intensity of neutron signals calculated over the series of succeeding time intervals after the shower passage. As it was discussed in Section~\ref{sectiista}, these series were originally recorded with a 250\,$\mu$s time resolution, and in strict synchronization with the EAS trigger signal. Since the applied gas discharge counters can detect charged particles of the shower front together with useful signal from neutron interactions, the amount of pulses registered in the leading---next-after-trigger---time interval in each event was somewhat corrected before averaging. This correction consists of a forced subtracting one unit from the number of pulses detected by each counter over duration of the leading interval which was made in all cases when this number occurred non-zero. After the correction, time distributions of neutron intensity detected in individual EASs were averaged between shower events which have close values of their size $N_e$ and of core distance $r$ from the point of neutron detector. For combination of the range limit conditions ($N_e=10^{6.0}-10^{6.5}$, $r\leqslant 10$\,m) the average distributions of such a kind are presented in Figure~\ref{figineutrotempo}.

Left plot in Figure~\ref{figineutrotempo} corresponds to the signals of all bare neutron counters in the point $N$ summed together, while the right one---to the counters which were operating within PVC moderator tubes with a 0.6~cm wall thickness. In both cases the total amount of detected neutron pulses was normalized to the number of counters in either group, and to the sensitive area of a single neutron counter.

Together with the intensity of the EAS connected signals, the corresponding background levels of the neutron counting rate are shown with thick dashed lines in both plots of Figure~\ref{figineutrotempo}. These levels were calculated over the data of a continuous monitoring type measurements which were going on uninterruptedly during all the time of the detector operation, as it was explained in Section~\ref{sectiista}. From their comparison it follows that an experimental relation between background intensities of free and moderator covered  counters is about ${R}\approx 1.6$, which is in agreement with the relation of the neutron detection efficiencies $\varepsilon$ resulting from the curves (\textit{1}) and (\textit{2}) in Figure~\ref{figieffici} for the thermal range of neutron energies ($E_n\approx 2\cdot 10^{-2}$\,eV). Indeed, it was quite natural to expect that the background flux should consist mostly of low-energy neutrons whose energies occur in temperature equilibrium with the outer environment. It is seen also in Figure~\ref{figineutrotempo} that the intensity of a surplus, EAS connected neutron flux for both counter groups evidently remains at a detectable level above its background at least up to the times of about 15--20\,ms after the passage of shower front.

\begin{figure*}
\begin{center}
\includegraphics[width=0.49\textwidth, trim=5mm 0mm 0mm 0mm]{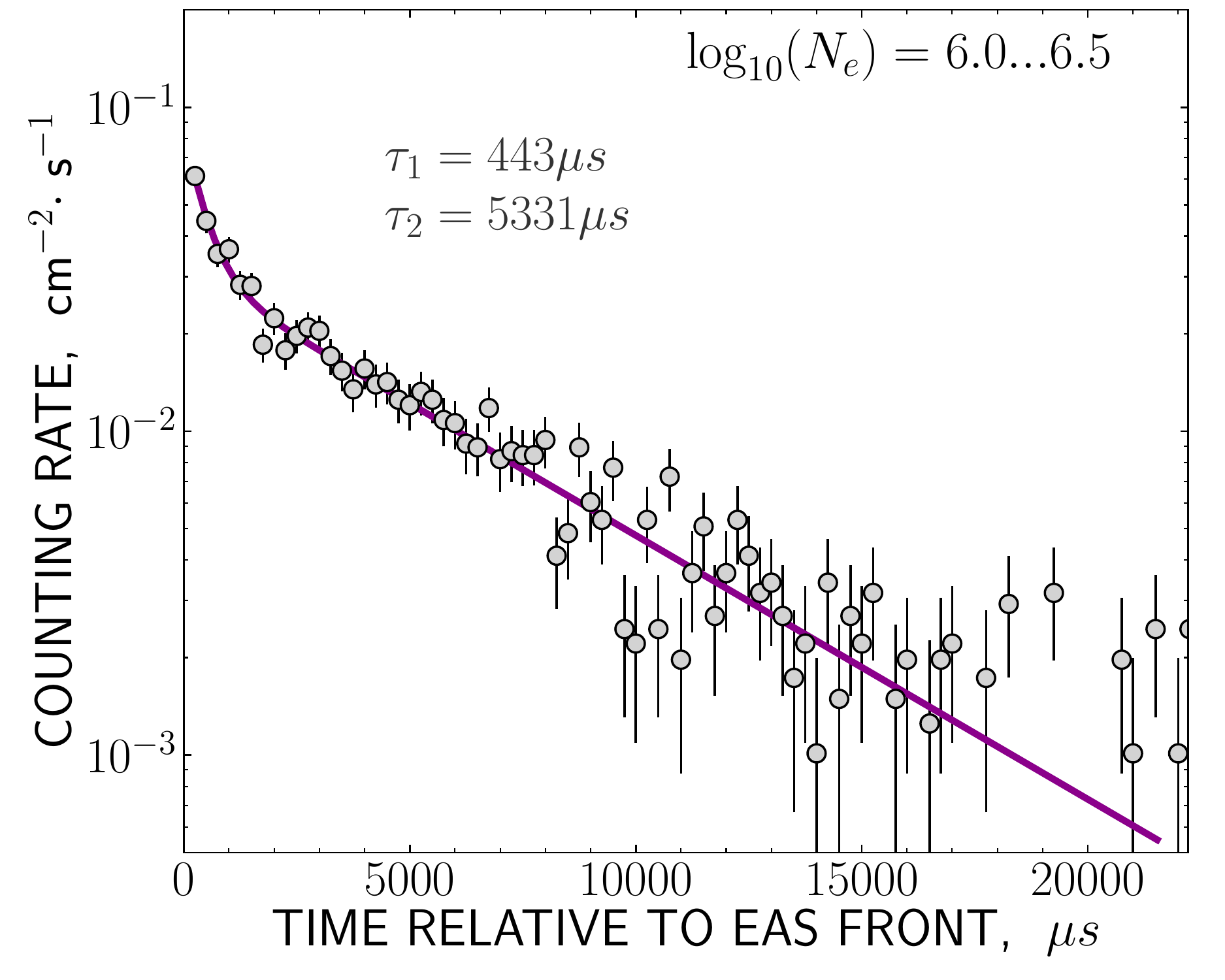}
\includegraphics[width=0.49\textwidth, trim=0mm 0mm 5mm 0mm]{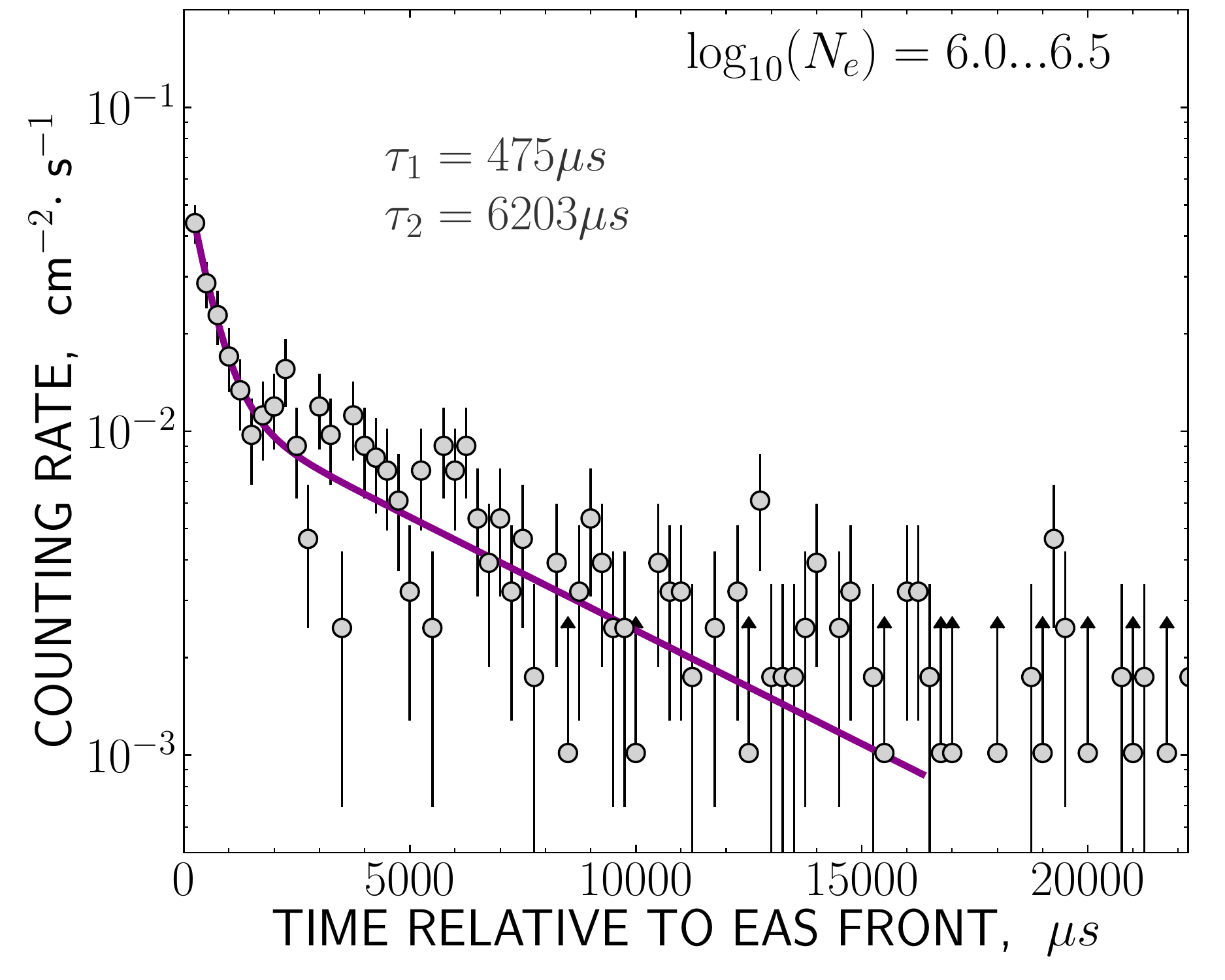}
\caption{Average distribution of the registered neutron intensity after EAS passage with subtraction of corresponding background levels for bare (left) and moderator coated (right) neutron counters. Smooth curves correspond to approximation of the experimental points with a sum of two exponents~(\ref{equaneutrotau}).}
\label{figineutrotempoappro}
\end{center}
\end{figure*}

In Figure~\ref{figineutrotempoappro} the same mean distributions of neutron counting rate measured with bare and moderator covered counters in central region of the $N_e=10^{6.0}-10^{6.5}$ EASs are presented in the form with a previously subtracted background. As it can be seen in such representation, for both counter groups the experimentally measured temporal behaviour of a purely EAS connected neutron deposit~$I(t)$  can be conveniently  approximated by a sum of two exponents,
\begin{equation}
I(t)=\sum_{i=0}^{i=2}a_i\exp(-t/\tau_i),
\label{equaneutrotau}
\end{equation}
with essentially different lifetime values: $\tau_1\approx 400-500$\,$\mu$s and $\tau_2\approx 5000-6000$\,$\mu$s. Similar results concerning the neutron accompaniment of the EAS were reported in \cite{stenkin-mephi2016,stenkin-yangbajing2016}; it was stated there that the exponential component with smaller $\tau$ corresponds to evaporation neutrons born by interaction of the EAS hadronic particles in the nearest vicinity to detector site, while the slowest exponent with a millisecond order lifetime indicates the arrival of neutrons from distant interactions at the periphery of the shower.

\begin{figure}
\begin{center}
\includegraphics[width=0.49\textwidth, trim=0mm 0mm 0mm 0mm]{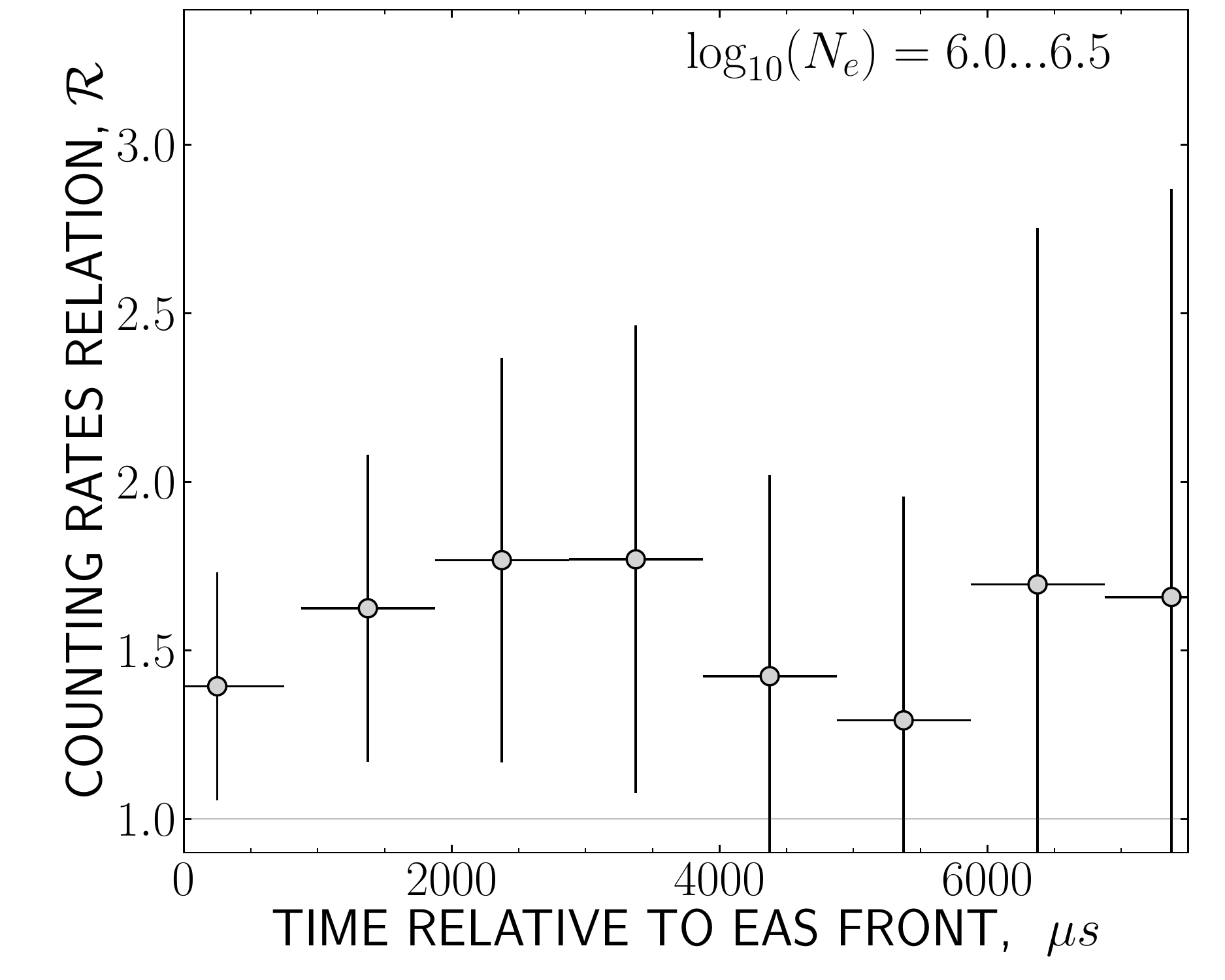}
% in fact, '..._55.pdf', not '..._60.pdf' - sic!
\caption{Dependence of the counting rate relation ${R}$ between the bare and moderator covered neutron detectors on the time since EAS passage.}
\label{figineutrorels}
\end{center}
\end{figure}

Pointwise dividing of average intensity curves from two plots of Figure~\ref{figineutrotempoappro} permits to trace the mutual ratio between momentary signal intensities of the bare and moderator covered neutron counters ${R}(t)=I_b(t)/I_m(t)$, as both were detected in the course of an EAS caused event development. Time dependence ${R}(t)$ obtained by such a way is presented in Figure~\ref{figineutrorels}, where it is seen that just at a delay time of $t\approx 100-200$\,$\mu$s, as well as afterwards, this relation is nearly constant and limited between the borders of $1.4-1.9$.  According to the neutron detection probability curves from Figure~\ref{figieffici}, observation of steady value $ {R}(t)>1$ means that since the very beginning of an event low-energy thermalized neutrons prevail among neutron flux connected with an EAS passage.

Temporal distributions of the intensity of neutron flux like the ones in Figures~\ref{figineutrotempo} and~\ref{figineutrotempoappro} can be of use when one attempts to define the total multiplicity $M$ of the neutrons born by the nuclear active shower particles in their interaction with the matter around the detector. Evidently, this multiplicity can be calculated by the integration of distribution curves over some fixed period---the \textit{gate} time ${T}_g$. This period is limited both at its lower side by the moment when charged shower particles of an EAS front were moving through the detector, and at its large side by the time when the exponentially diminishing intensity of EAS connected neutrons falls down to the background level.

Expanding the gate $ {T}_g$ into a short time range (below a few microseconds) leads to unwanted excess of the sum pulse count which is due to the signals from charged particles of the shower front. As it was just mentioned above, a proper correction of this effect consists in forced subtraction of one unit from signal multiplicities which have been detected by every counter in the course of the leading (the first-after-trigger) time interval. With account to typical duration of the shaped pulse signals at the output of neutron counter channel, such subtraction means that the actual lower limit of the gate time $T_g^{low}$ is about 3--5\,$\mu$s.

Too long duration of the gate time can cause excessive\  \ registration\ \ of\ \ background\ \ neutrons which have no connection with considered shower event. Taking into account a specific exponential decay shape of the neutron counting rate dependence $I(t)$ after the shower passage, one can conclude that the proper upper limit for $ {T}_g$ must be comparable with the biggest lifetime value $\tau_2$ in formula~(\ref{equaneutrotau}). On the assumption of concrete technical condition---fixed 250\,$\mu$s length of an elementary time interval, it was decided to accept the upper gate limit as $T_g^{upp}=8500$\,$\mu$s which corresponds to $\approx$$1.5\cdot\tau_2$. This seems to be a satisfactory compromise between sufficiently prolonged duration of the signal collection time in the case of large multiplicity neutron events and tolerable contamination of the detected  multiplicity of useful signals on the part of background neutron flux in the opposite case. Indeed, with the level of background counting rate intensity shown in Figure~\ref{figineutrotempo} the number of random pulses which can hit $T_g=8500$\,$\mu$s time lapse is $\approx$$0.02$ per centimeter square (for bare neutron counters), while the integration of the corresponding exponential curve $I(t)$ from Figure~\ref{figineutrotempoappro} within the limits of $T_g^{low}$\ldots$T_g^{upp}$ gives the value of $0.16$\,cm$^{-2}$. Hence, in the case of exemplary counting rate distribution for the $N_e=10^{6.0}-10^{6.5}$~ shower size interval relative share of random background signals with accepted $T_g$ is $0.02 / 0.16 \times 100 \approx$$12$\%. For $N_e=10^{5.0}-10^{5.5}$~size interval this estimation occurs to be $28$\%, and for the EAS events with $N_e\sim 10^{7.0}$---of about 2\% only.
%import numpy, scipy.integrate ;
%from r09008cls import APPRO_EXPO2 ;

%LEFT = 5. ;       # both
%RIGHT = 8500. ; # in us.

%#PARAMS = ( 0.05458683,0.00226862,0.03119264,0.00018805 ) ;
%  # approximation for 'without moder' case and $\log_{10}(N_e)=6.0...6.5$
%PARAMS = ( 0.05119596, 0.00153374, 0.02770687, 0.00012053 ) ;
%  # background is not subtracted

%#PARAMS = ( 0.04969643, 0.00607745, 0.00444587, 0.00020548 ) ;
%  # approximation for 'without moder' case and $\log_{10}(N_e)=5.0...5.5$
%  # ( points: 11 ... 70, inits: 10,0.01,0.1,0.001 ).
%#PARAMS = ( 0.00825,0.00226862,0.00465,0.00018805 ) ;
%#   0.15*APPRO_EXPO2( x, [ 0.05458683, 0.00226862, 0.03119264, 0.00018805 ] )
%#PARAMS = ( 0.00422409654, 0.000328806733, 0.0040560775, 1.47867003e-05 ) ;

%#PARAMS = ( 0.19265421, 0.00032395, 0.25880229, 0.00032395 ) ;
%  # approximation for 'without moder' case and $\log_{10}(N_e)=7.0...7.5$
%  # ( points: 10 ... 50, inits: 1,0.01,0.1,0.001 ).
%PARAMS = ( 0.41, 0.078297954, 0.25, 0.000266008366 ) ;
%# 8*APPRO_EXPO2( x, [ 0.05119596, 0.00153374, 0.02770687, 0.00012053 ] )
%
%
%p = scipy.integrate.quad( lambda x : APPRO_EXPO2( x, PARAMS ), \
%                          LEFT, RIGHT, full_output=1 )[ : 2 ] ;
%q = scipy.integrate.quad( lambda x : 0.0025, \
%                          LEFT, RIGHT, full_output=1 )[ : 2 ] ;
%
%print LEFT, RIGHT, p, q ;
%print '-->',  q[ 0 ] / p[ 0 ] * 100, '%' ;

A maximum error value which could arise because of the limited gate time duration can be estimated through the integration of exponential approximation $I(t)$ of experimental intensity curves. With above $T_g$ limits and lifetime estimations $\tau_1\approx 450$\,$\mu$s and $\tau_2\approx 5500$\,$\mu$s which were mentioned by discussion of the plots in Figure~\ref{figineutrotempoappro}, this error occurs to be of about 20\%:
\begin{equation}
1 - \biggm(\int_{5}^{8500}I(t)dt \biggm/ \int_0^\infty I(t)dt\biggm) \approx 0.18.
\label{equaneutrotauerro}
\end{equation}
% limited_gate_time_error_script
%import sys, numpy, scipy.integrate ;
%from r09008cls import APPRO_EXPO2 ;

%LEFT = 0. ;       # both
%RIGHT = 100000. ; # in us.
%
%PARAMS = ( 0.05458683,0.00226862,0.03119264,0.00018805 ) ;
%    # approximation for 'without moder' case.
%
%if len( sys.argv ) > 1 :
%  RIGHT = float( sys.argv[ 1 ] ) ;
%if len( sys.argv ) > 2 :
%  LEFT = float( sys.argv[ 2 ] ) ;
%   # 'right' is BEFORE 'left' - sic!

%x = numpy.arange( LEFT, RIGHT, LEFT + ( RIGHT - LEFT ) / 10000. ) ;
%print x
%print APPRO_EXPO2( x, PARAMS ) ;

%print LEFT, RIGHT, \
%  scipy.integrate.quad( lambda x : APPRO_EXPO2( x, PARAMS ), \
%                          LEFT, RIGHT, full_output=1 )[ : 2 ] ;
%
% integral for the limits    0...\infty: 190.
% integral for the limits    0...5:      0.43.
% integral for the limits    5...8500:    156.
% integral for the limits 8500...100000:   34.
% relation: (0.43+34)/190.= 0.22.

All the results presented hereafter on the multiplicity of EAS connected neutron accompaniment were obtained with these limits of the gate time $ {T}_g$.

\subsection{The multiplicity of neutron signals}
\label{sectineutromulti}

\begin{figure}
\begin{center}
\includegraphics[width=0.5\textwidth, trim=14mm 0mm 2mm 5mm]{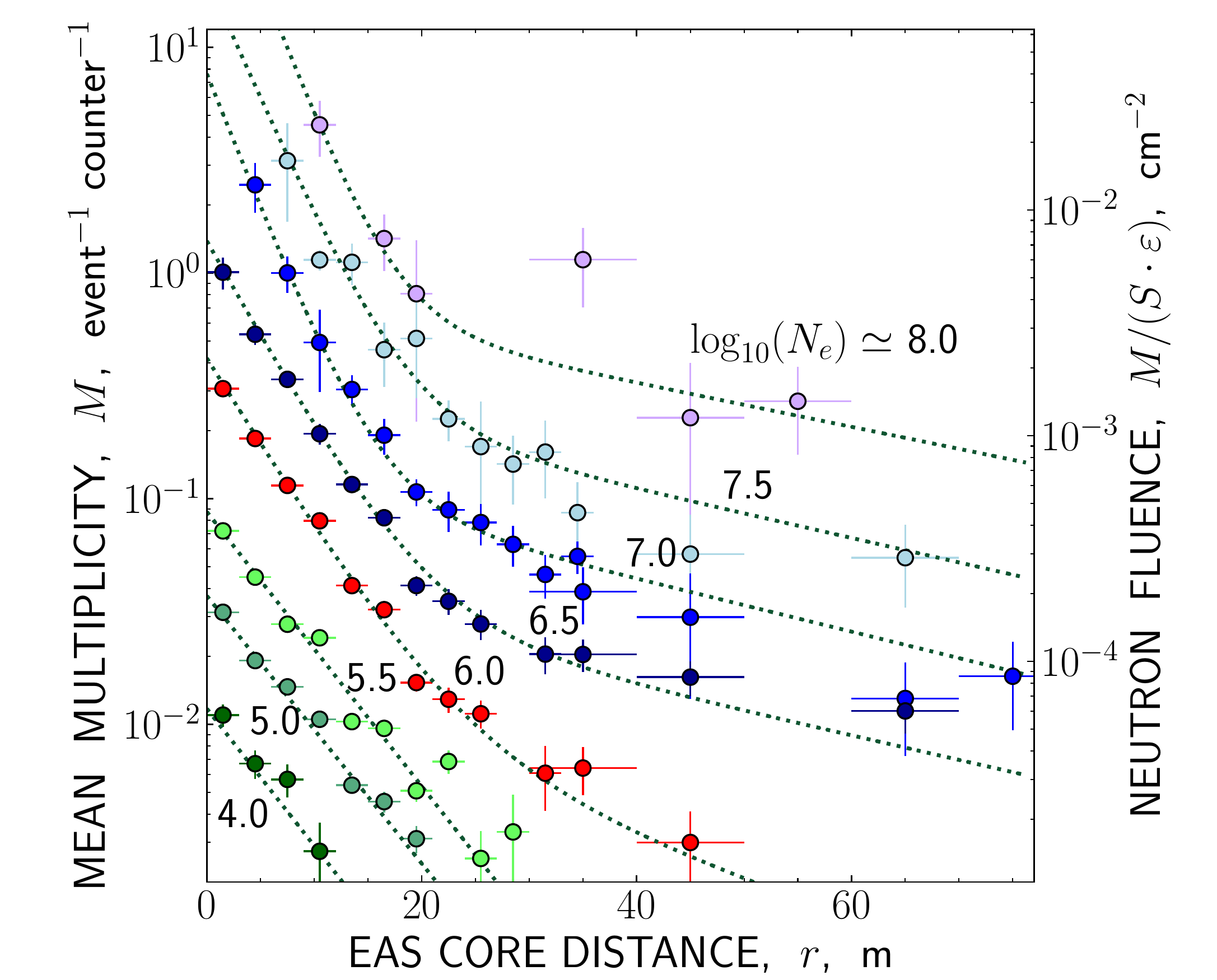}
\caption{Spatial distribution of the average multiplicity of neutron signals $M$ and corresponding local fluence of EAS connected neutrons for various ranges of the shower size $N_e$. Points are experimental data, dotted lines mark their exponential fit $D(r)$ (see text), and the numbers beside curves mean the decimal logarithm of the average size parameter of the corresponding EAS. The gate time of the neutron signal collection after an EAS passage is 8500\,$\mu$s, the background is subtracted.}
\label{figineutrofpr}
\end{center}
\end{figure}

In experimental practice, the total fluence of an extensive air shower connected neutrons which result from the interaction of the hadronic EAS component in the outer environment can be calculated as
\begin{equation}
F_n=M/(S\cdot \varepsilon),
\label{equoflux}
\end{equation}
where the multiplicity $M$ is the sum number of the pulse signals which have come from the counters of a given neutron detector over the duration of the gate time $ {T}_g$, $S$ is the sum sensitive area of the counters installed in that detector point, and $\varepsilon$---the probability of neutron registration. (As it was shown in previous section, the EAS accompaniment consists mostly of neutrons which belong to the thermal energy range; hence, for example in the case of a bare neutron counter the curve \textit{(1)} in Figure~\ref{figieffici} gives the probability $\varepsilon\approx 0.2$). In turn, grouping the showers by their size $N_e$ and the distance $r$ between the shower axis and the location of neutron detector allows to analyze spatial distribution of the average multiplicity and fluence values, as well as their dependence on the size of the corresponding shower. Such a kind of average distributions for all EAS events which were detected so far at the Tien Shan mountain station is presented in Figure~\ref{figineutrofpr}.

Each experimental data point in the plot of this figure was obtained by the following procedure. For every registered EAS it was defined a set of shower parameters including the shower size $N_e$ and the position of its axis, then distances $r$ were calculated between the shower center and the location places of all three neutron detectors. Further on, the multiplicities $M$ of neutron signals which fall into $T_g$ period after this EAS were defined and averaged between the cases with close combination of the $N_e$ and $r$ parameters, independently for each detector point. After normalization to the amount of neutron counters installed in respective point and subtraction of the background intensity level, the values thus calculated were plotted in Figure~\ref{figineutrofpr} as a set of average spatial distributions of the multiplicity of neutron signals $M$ in EAS events which belong to different ranges of $N_e$. Then the corresponding mean values of the local neutron fluence can be estimated by the auxiliary right axis of this plot which is graduated in accordance with the above mentioned formula (\ref{equoflux}).

According to the data points in Figure~\ref{figineutrofpr}, generally the mean multiplicity of neutron accompaniment diminishes ra\-pid\-ly with the distance $r$ to EAS center, so all distributions in this plot can be conveniently approximated by a family of exponential functions:
\begin{equation}
D(r)=\sum_i a_i\exp(-r/\rho_i).
\label{equarho}
\end{equation}
The spatial scale parameters $\rho_i$ in these approximations were defined through minimizing of a $\chi^2$ like discrepancy sum between experimental $M$ points and corresponding $D(r)$ values which was made individually for every $N_e$ range distribution in Figure~\ref{figineutrofpr}. The resulting best fit curves $D(r)$ are shown there with dotted lines, while the corresponding $\rho$ values are listed in Table~\ref{tabneutrorhos}.

\begin{table}
\begin{center}
\caption{Best fit parameters of exponential  functions $D(r)$ which approximate experimental distributions of neutron multiplicity in Figure~\ref{figineutrofpr}.}
\label{tabneutrorhos}
%\begin{tabular*}{\columnwidth}{@{\extracolsep{\fill}}ccc@{}}
\begin{tabular*}{\columnwidth}{@{\extracolsep{0.25\columnwidth}}ccc}
\hline
$\log_{10}N_e$ & $\rho_1, m$ & $\rho_2$, m \\
\hline
% #1
$4.0\ldots 4.5$ &
7.1 &
--- \\

%#2
$4.5\ldots 5.0$&
7.0&
--- \\

% #3
$5.0\ldots 5.5$&
7.4&
--- \\

% #4
$5.5\ldots 6.0$&
7.1&
--- \\

% #5
$6.0\ldots 6.5$&
5.6&
25 \\

% #6
$6.5\ldots 7.0$&
5.0&
40 \\

% #7
$7.0\ldots 7.5$&
3.6&
37 \\

% #8
$7.5\ldots 8.0$&
4.0&
40 \\

% #9
$\simeq 8.0$&
3.5&
44 \\

\hline
\end{tabular*}
\end{center}
\end{table}

As it follows both from Figure~\ref{figineutrofpr} and Table~\ref{tabneutrorhos}, spatial distributions of the EAS connected neutron multiplicity can be divided into two groups. First, there are low-intensity, single-exponent distributions commonly met in small shower events with the size $N_e\lesssim 10^6$. Typical value of the spatial $\rho$ parameter in these distributions is below 10\,m, so the neutron multiplicity does decrease very quickly with the rise of the shower core distance. On the contrary, in $D(r)$ distributions which belong to the shower events with $N_e\gtrsim 10^6$ together with this ``narrow'' component it appears another exponent with characteristic $\rho\approx 20-40$\,m. Due to this ``wide spread'' deposit, in the large size EAS a noticeable neutron flux can be still detected at rather big distances, at least of some tens of meters order from the shower center.

\begin{figure}
\begin{center}
\includegraphics[width=0.49\textwidth, trim=0mm 0mm 15mm 0mm]{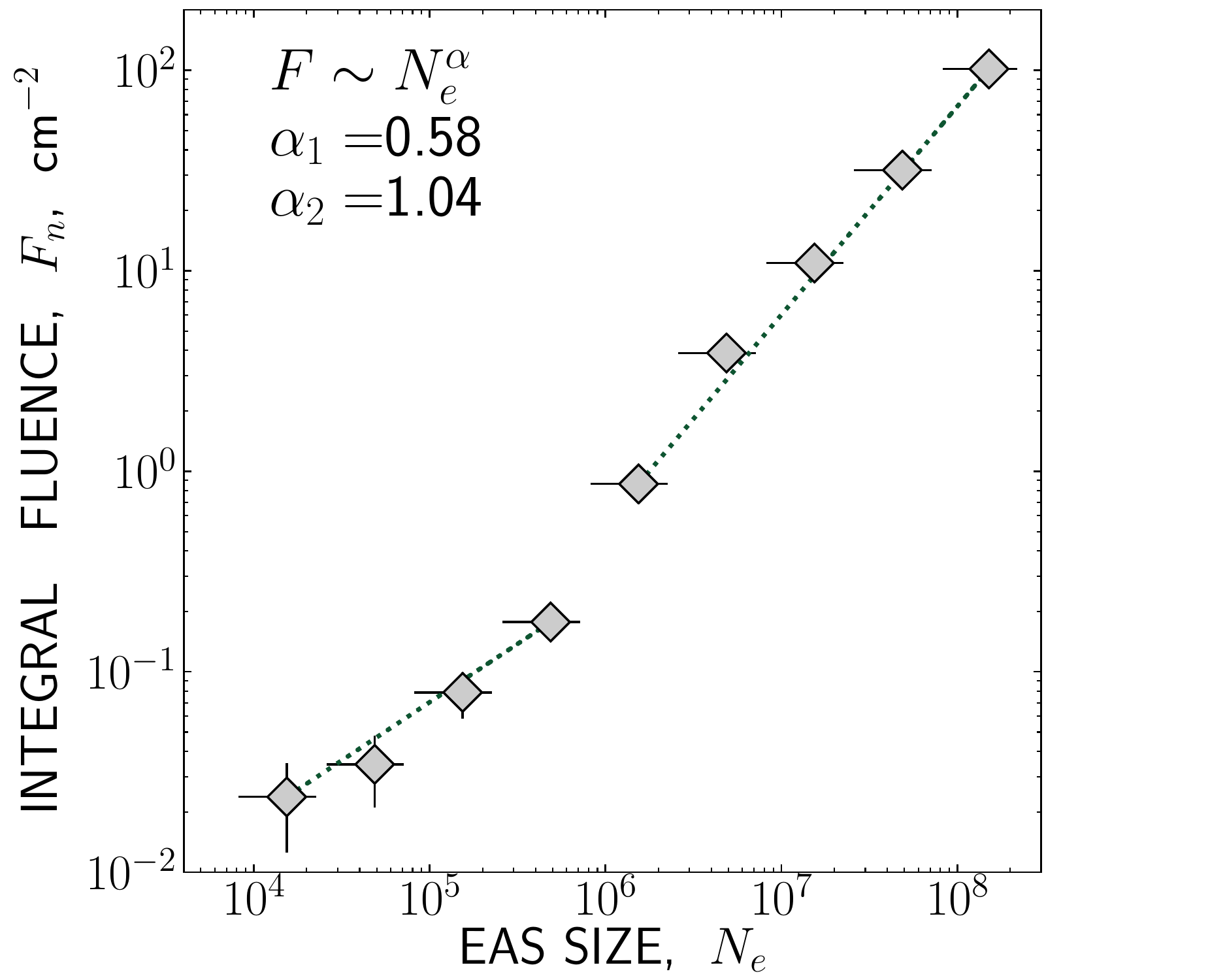}
\caption{Average neutron fluence $F_n$ in dependence on the size of corresponding extensive air shower $N_e$. Dotted lines mark the result of a least square approximation of the experimental points with a piecewise power function.}
\label{figineutromulti}
\end{center}
\end{figure}

Numerical integration of the spatial distribution functions $D(r)$ obtained for various ranges of shower sizes gives average integral fluence of the whole neutron flux which should be expected after the passage of an EAS with given $N_e$:
\begin{equation}
F_n=\int_0^\infty D(r)\cdot 2\pi r dr \biggm / ( S \cdot \varepsilon)
\label{equaoitegrho}
\end{equation}
(as before, $S$ here is the sensitive area of a single neutron counter, and $\varepsilon$---its efficiency in the range of thermal neutron energies). These values, after additional introducing a 20\% correction (\ref{equaneutrotauerro}) for a limited duration of the neutron collection gate time are presented in Figure~\ref{figineutromulti}. As it is seen there, generally the dependence of the integral neutron fluence $F_n$ on the shower size $N_e$ follows to a piecewise power function $F_n\sim N_e^\alpha$ which experiences a sharp change of its power index around $N_e\approx 10^6$: below this border limit the value $\alpha_1$ is of about 0.6, while above it a least square approximation of the experimental data points in Figure~\ref{figineutromulti} gives $\alpha_2\approx 1.0$.

\subsection{The multiplicity of low-energy neutrons and the~problem of the~knee in the~primary cosmic ray spectrum}
\label{sectineutromultiknee}

Observation of any nonuniformity in multiplicity behaviour of the neutron accompaniment of EAS, like the one in Figure~\ref{figineutromulti} means the existence of some corresponding change in average characteristics of the hadronic component of extensive air showers, which is the original ancestor of detected neutrons. On the other hand, it is well known that the mean number of evaporation neutrons produced in a nuclear reaction depends on the interaction energy, \textit{i.\,e.} on the energy $E_h$ of an incident hadron (generally, this dependence is of a power like shape, $M\sim E_h^\gamma$ with $\gamma\approx 0.3-0.5$), while the total amount of such interactions evidently must be proportional to the full number of hadrons in a shower ($M\sim N_h$). As a consequence, a sudden change in power index of the $F_n(N_e)$ dependence in Figure~\ref{figineutromulti} means the existence of some additional increase in the mean energy $\overline{E}_h$ of EAS had\-ro\-nic particles, or in the average hadron number $\overline{N}_h$ in EAS, or in both, and this inequality necessarily takes place at the $N_e\approx 10^6$ threshold value. Of these alternatives a rise of $\overline{N}_h$ in the large-size EAS seems to be most probable, since the $\gamma$ parameter in the mentioned $M(E_h)$ dependence is essentially below one unit, so any enhancement in the mean hadron energy occurs less effective from the viewpoint of neutron production. Hence, a mechanism with generation of a large amount of hadrons is more preferable for explanation of multiplicity features of the detected neutron flux in shower events.

Distinct localization of the lateral distribution functions in Figure~\ref{figineutrofpr} in a close neighbourhood of shower axis, that is just within that spatial region where the EAS hadrons do mostly concentrate is another argument in favor of the above considerations on the possible origin of the detected neutron signal and the cause of its peculiarities.

It is important to mention that in accordance with standard formula which was used for recalculation between $N_e$ and $E_0$ at the time of former shower experiments at Tien Shan, the range of EAS sizes $N_e\approx 10^6$ corresponds to the primary particle energy $E_0\approx 3\cdot 10^{15}$~eV \cite{ontien_icrc1987__e0_through_ne_ru,hadron_spc_2017}, that is to the position of the well-known knee in the energy spectrum of primary cosmic rays. The fact that the above described changes both in the shape of  lateral distribution of the EAS accompanying neutron flux, and in the dependency of its integral fluence on shower size occur just at this same threshold point on the energy scale permits to add these effects to a large list of various peculiar phenomena found in this energy range, and particularly in experiments which were held previously at the Tien Shan mountain station \cite{ontien-nim2016}.

It should also be noted that the above explanation of the presented experimental results means a sudden opening in the $N_e\approx 10^6$ EASs range of some additional channel for effective energy transmission into a multitude of (possibly, low-energy) hadrons concentrated within a rather tight spatial region of shower core. Such an EAS component could hardly be registered with usual high threshold detectors of a typically wide spread shower installation, while its deposit in the whole energy balance of the shower might be quite noticeable. As a consequence, such an effect could be a plausible failure reason of many explanation attempts which were given until now to the origin of the $E_0\approx 3\cdot 10^{15}$~eV knee in the primary cosmic ray spectrum.

\section{Post-EAS gamma radiation}

\subsection{Low-energy gamma ray accompaniment within the central region of a large size EAS}

At the time when the sample shower events from Figure~\ref{figineutroevents} were obtained, low-energy gamma ray detector was operating together with the neutron one, both being placed in vicinity to the center point of the ``carpet'' of EAS particles detectors. Hence, it was possible to trace the temporal behaviour of soft gamma radiation together with that of the neutron signal in the detected shower events. By these measurements the same 250\,$\mu$s time granularity of the input pulse intensity recording, and the same whole second long duration of the signal collection window were applied as in the case of the neutron counters.

\begin{figure*}
\begin{center}
\includegraphics[width=0.49\textwidth, trim=11mm 22mm 66mm 0mm]{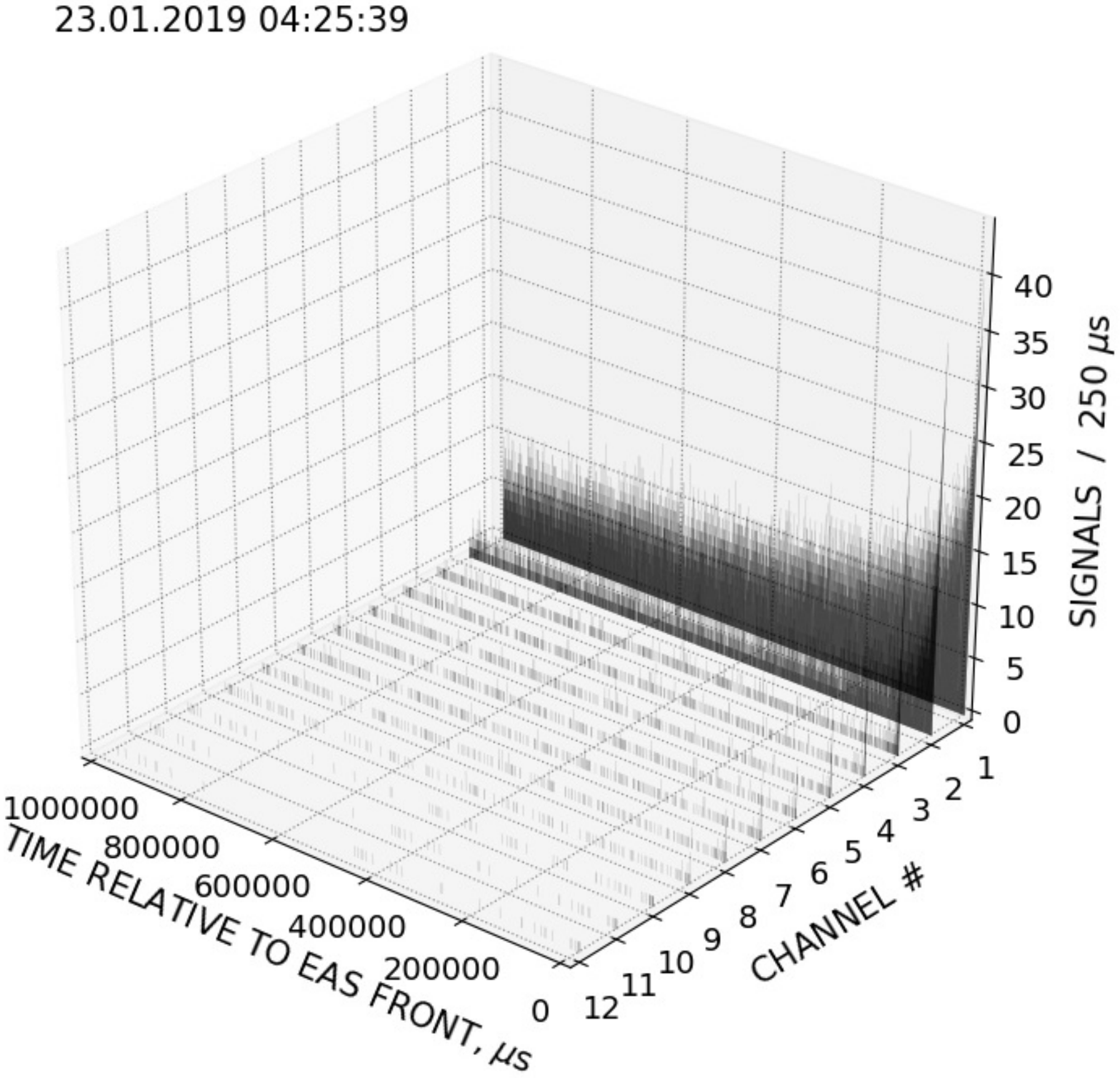}
\includegraphics[width=0.47\textwidth, trim=0mm 0mm 0mm 0mm]{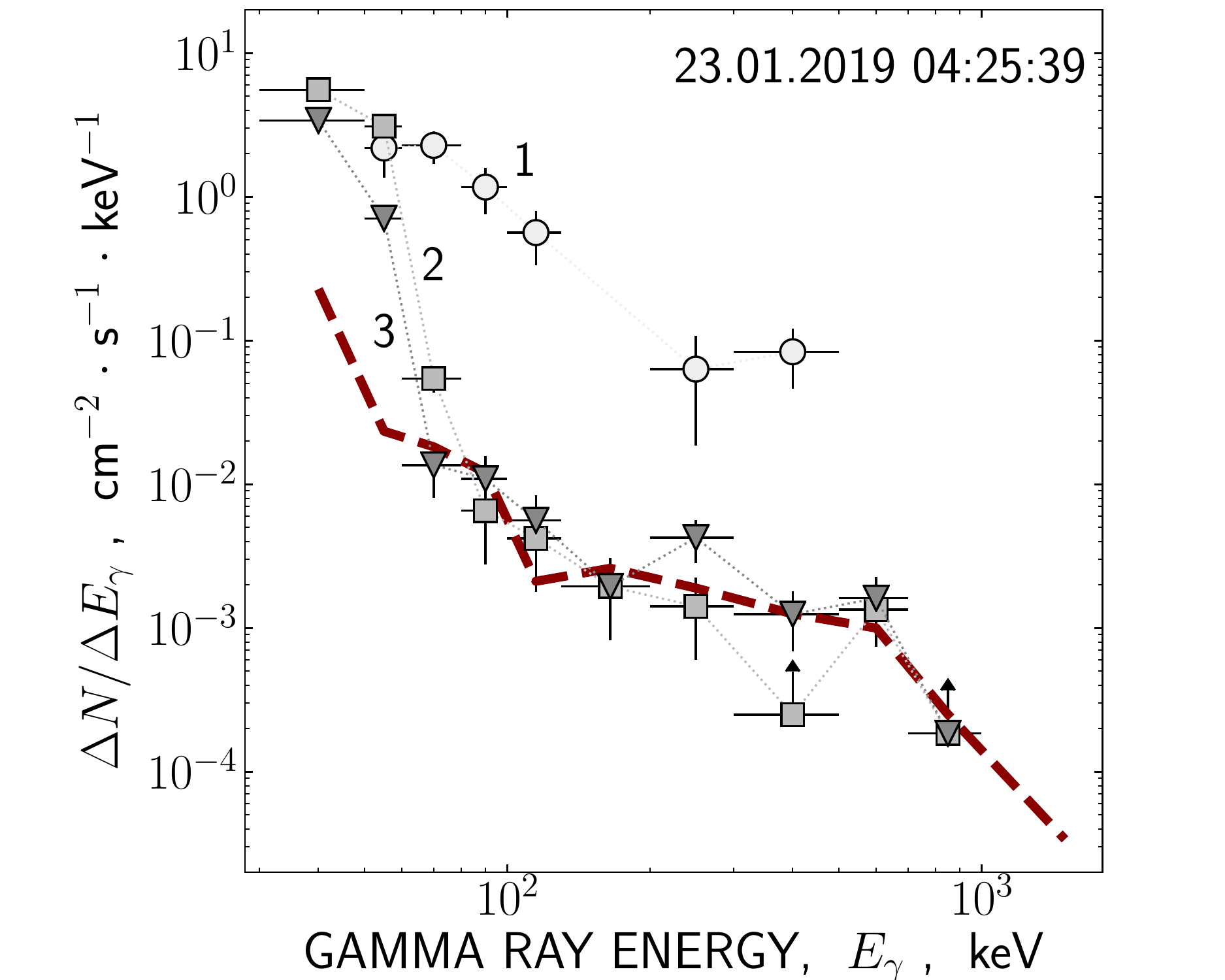}
\\
\includegraphics[width=0.49\textwidth, trim=11mm 22mm 66mm 0mm]{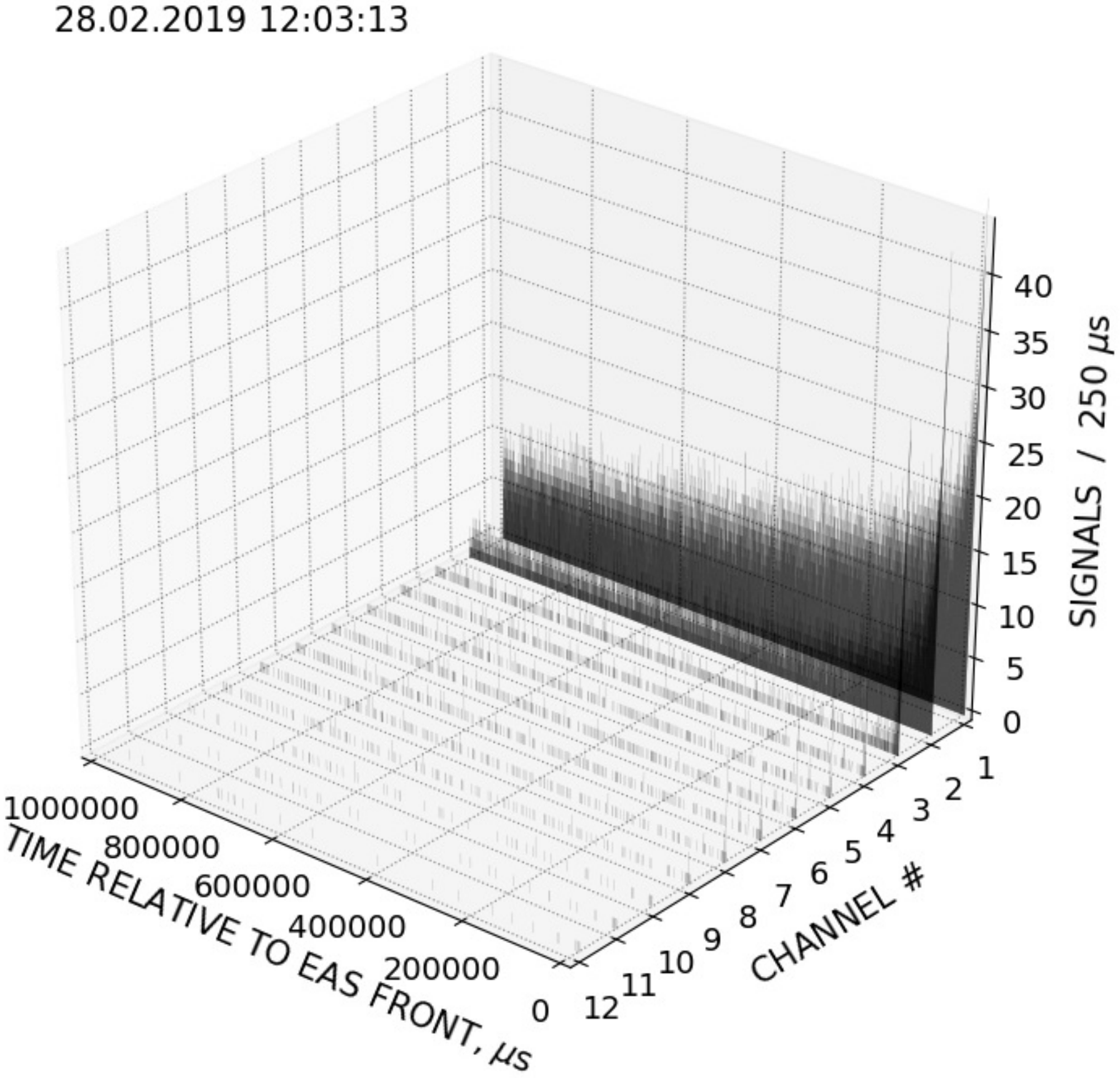}
\includegraphics[width=0.47\textwidth, trim=0mm 0mm 0mm 0mm]{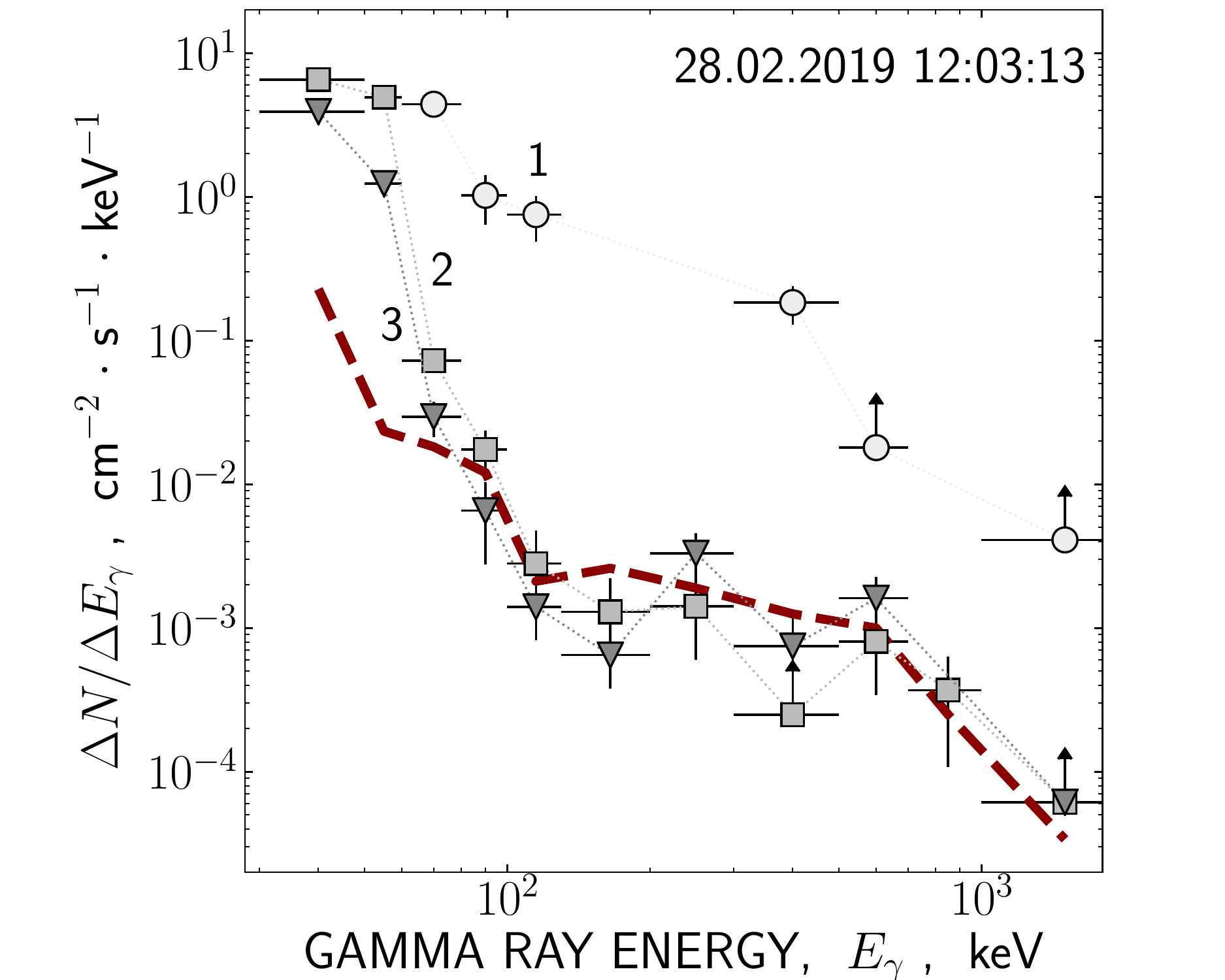}
\\
\includegraphics[width=0.49\textwidth, trim=11mm 22mm 66mm 0mm]{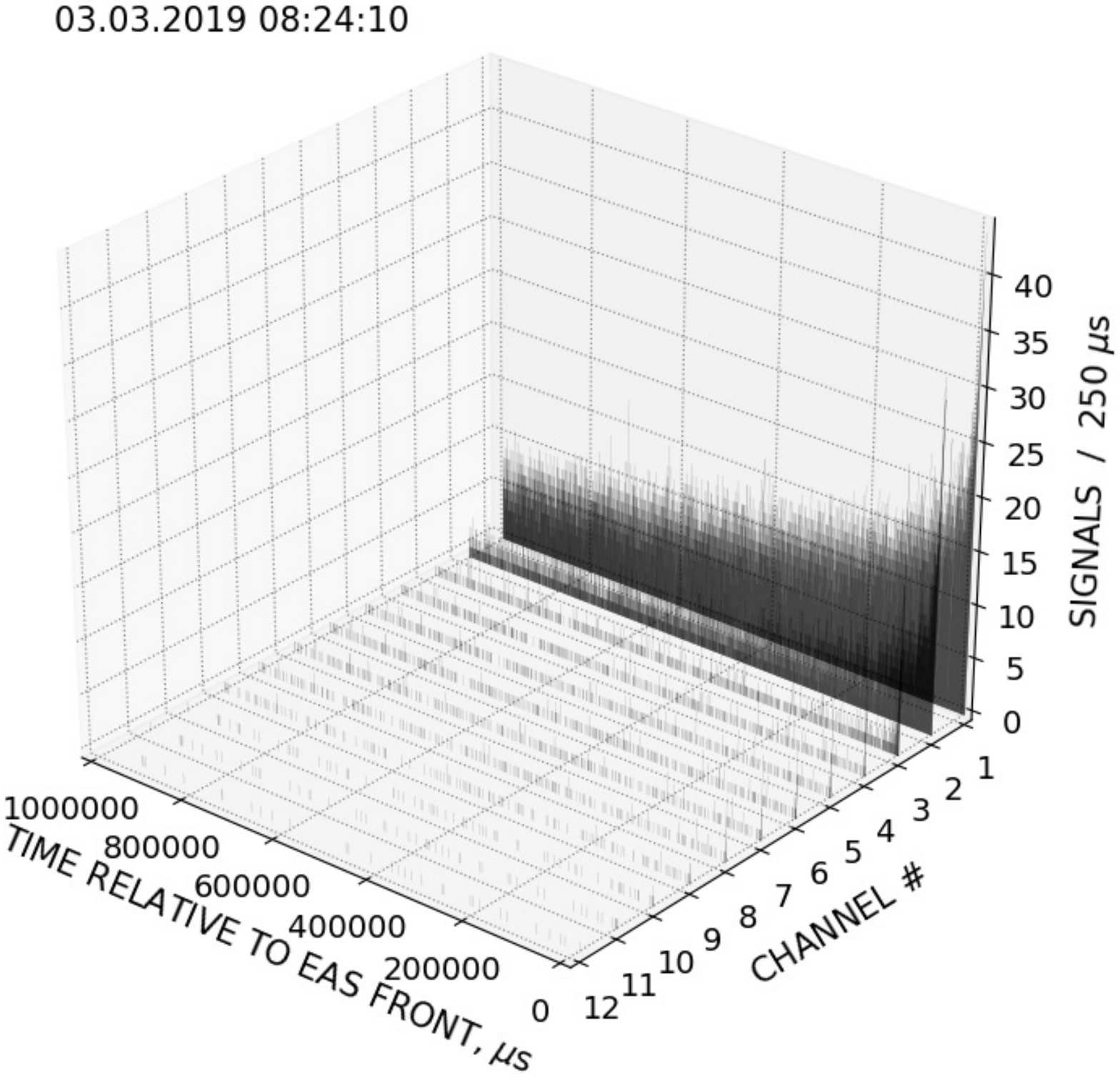}
\includegraphics[width=0.47\textwidth, trim=0mm 0mm 0mm 0mm]{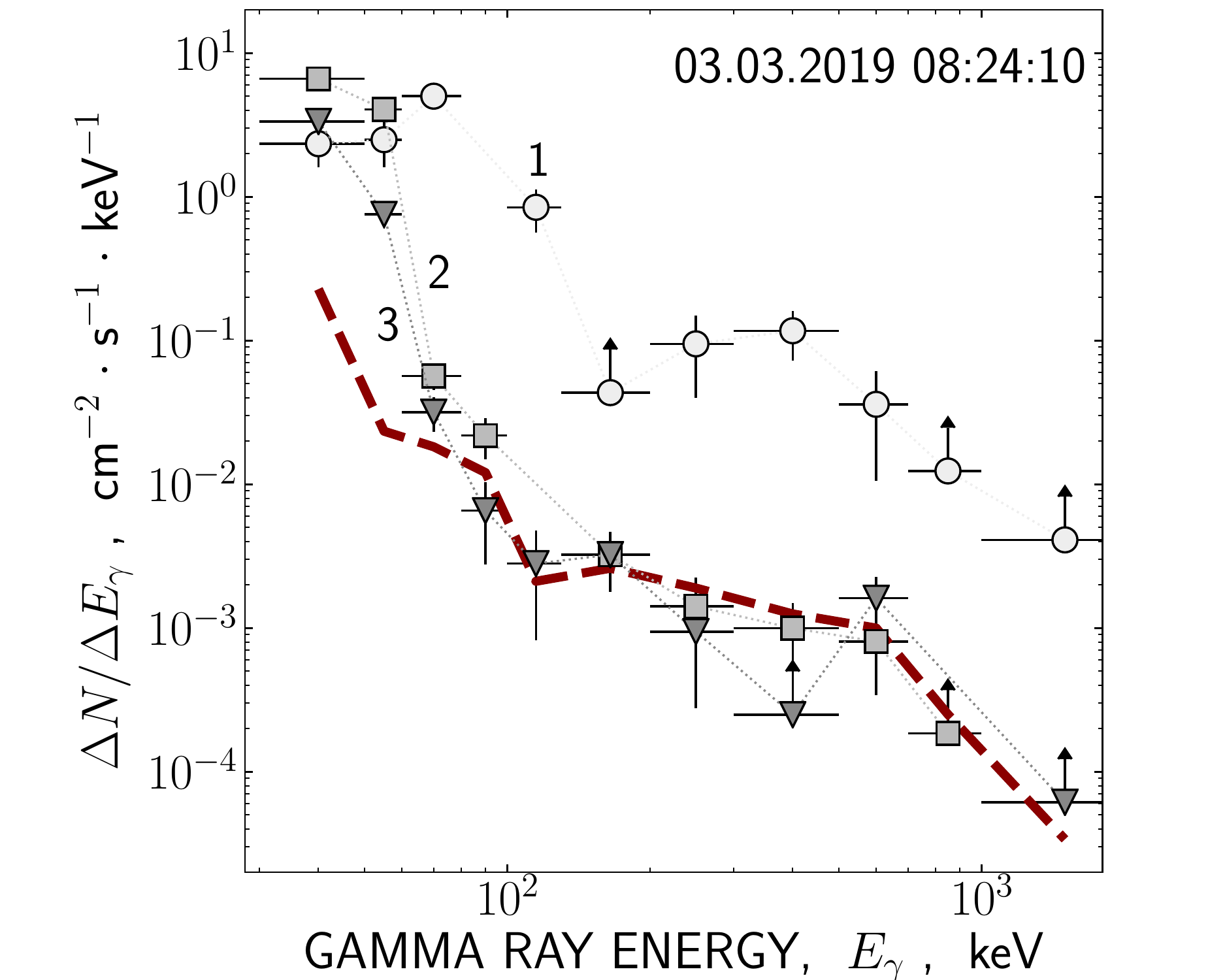}
\caption{Time distributions (in the left column) of output signals which were detected in 12 amplitude channels of gamma ray detector in the large size EAS events from Figure~\ref{figineutroevents}, and corresponding differential energy spectra of gamma radiation calculated over the time spans of 0\,\ldots 0.5 (\textit{1}), 100\,\ldots 150 (\textit{2}), and 900\,\ldots 950 (\textit{3}) milliseconds after shower passage in these events. Dashed lines mark the spectrum of background radiation.}
\label{figigammoevents}
\end{center}
\end{figure*}

For a set of the large size EASs considered above in Section~\ref{sectineutrotempo} the intensity distribution of delayed gamma ray signals over the time since the passage of shower front are presented in the left column plots of Figure~\ref{figigammoevents}. These plots correspond to the data from the gamma detector which was installed in the point designated as $N$ in Figure~\ref{figineutroevents}, and thus has occurred within the central EAS region in all considered events. Output pulse signals from this detector were connected to a dozen amplitude discriminators with a stepwise increasing thresholds, so it was possible to estimate integral energy spectrum of gamma radiation immediately during the measurements. Correspondingly, in the left column plots of Figure~\ref{figigammoevents} temporal distributions of the signals detected in successive amplitude channels are presented by a number of different time series each of which relates to a particular channel, while the equivalent energy thresholds are listed in Table~\ref{tabgammothreshs}. Besides these thresholds, in the rows of this table it is given the estimation of gamma ray detection probability (detector efficiency), as it follows for proper $E_\gamma$ value from second plot in Figure~\ref{figieffici}, and the intensity of background counting rate defined experimentally in  monitoring type measurements.

The right column plots in Figure~\ref{figigammoevents} present a number of differential energy spectra for considered EAS events which were calculated according to the signal counting rates in successive amplitude channels of the gamma ray detector. This calculation was made   with account to the full sensitive area of scintillation crystal, and to the probability of radiation detection in corresponding energy range, as the latter was defined in the ``efficiency'' column of Table~\ref{tabgammothreshs}. In each event the spectra were built separately for a number of different time delays in relation to EAS: over the times period of 0\,\ldots 0.5\,ms (\textit{i.\,e.}\,\,shortly after the passage of a shower front), 100\,\ldots 150\,ms, and 900\,\ldots 950\,ms. With  continuous dashed line in all spectra plots of Figure~\ref{figigammoevents} it is designated the spectrum of the background gamma radiation, as it results from the data listed in the ``background'' column of Table~\ref{tabgammothreshs}.

\begin{table*}
\caption{Energy thresholds set in twelve amplitude channels of gamma ray detector, the corresponding detection efficiency, and registered intensity of the background counting rate in successive channels.}
\label{tabgammothreshs}
\begin{center}
\begin{tabular*}{1.5\columnwidth}{@{\extracolsep{ 1ex } } cccc }
\hline
channel \#&
threshold, $E_\gamma$, keV&
efficiency, $\varepsilon$&
\parbox[c][8ex]{ 0.4\columnwidth }{\centering
background, $R_{bckgr}(\geqslant E_\gamma)$, cm$^{-2} \cdot$\,s$^{-1}$ } \\
\hline

1&
30&
0.50&
7.4 \\

2&
50&
0.75&
1.9 \\

3&
60&
0.77&
1.6 \\

4&
80&
0.80&
1.2 \\

5&
100&
0.83&
0.90 \\

6&
150&
0.77&
0.88 \\

7&
200&
0.74&
0.76 \\

8&
300&
0.70&
0.60 \\

9&
500&
0.65&
0.38 \\

10&
700&
0.60&
0.18 \\

11&
1000&
0.57&
0.12 \\

12&
2000&
0.50 &
0.10 \\

\hline
\end{tabular*}
\end{center}
\end{table*}

The most remarkable common feature in the presented EAS events is  an extraordinarily prolonged succession of output pulses which was observed in the channels with lowest amplitude threshold of the gamma  detector. As it is seen in the left column plots of Figure~\ref{figigammoevents}, in the range of radiation energies $E_\gamma \leqslant 50$\,keV such a delayed signal remains existing up to a whole second order times since the EAS passage, and evidently continues even later, with a very slow tendency to complete extinction. Considering the spectra plots in Figure~\ref{figigammoevents}, one can state that in the range of gamma ray energies above a few tens of keV any noticeable gamma ray flux above background can be detected only if a delay time after the shower does not exceed some hundreds of microseconds (these are spectra (\textit{1}) in this figure); later on its intensity falls down to the background level. Thus, it can be stated that the  radiation signal with a few seconds order lifetime  observed in central EAS region is due exclusively to the low-energy gamma rays with $E_\gamma \approx 30-50$\,keV.

Observation of a such unexpectedly prolonged flux of soft radiation in Tien Shan cosmic ray experiment was a motive for further investigation of the properties of EAS connected gamma ray accompaniment.

\subsection{Average time distributions of gamma ray intensity}

\begin{figure*}
\begin{center}
\includegraphics[width=0.49\textwidth, trim=3mm 0mm 3mm 0mm]{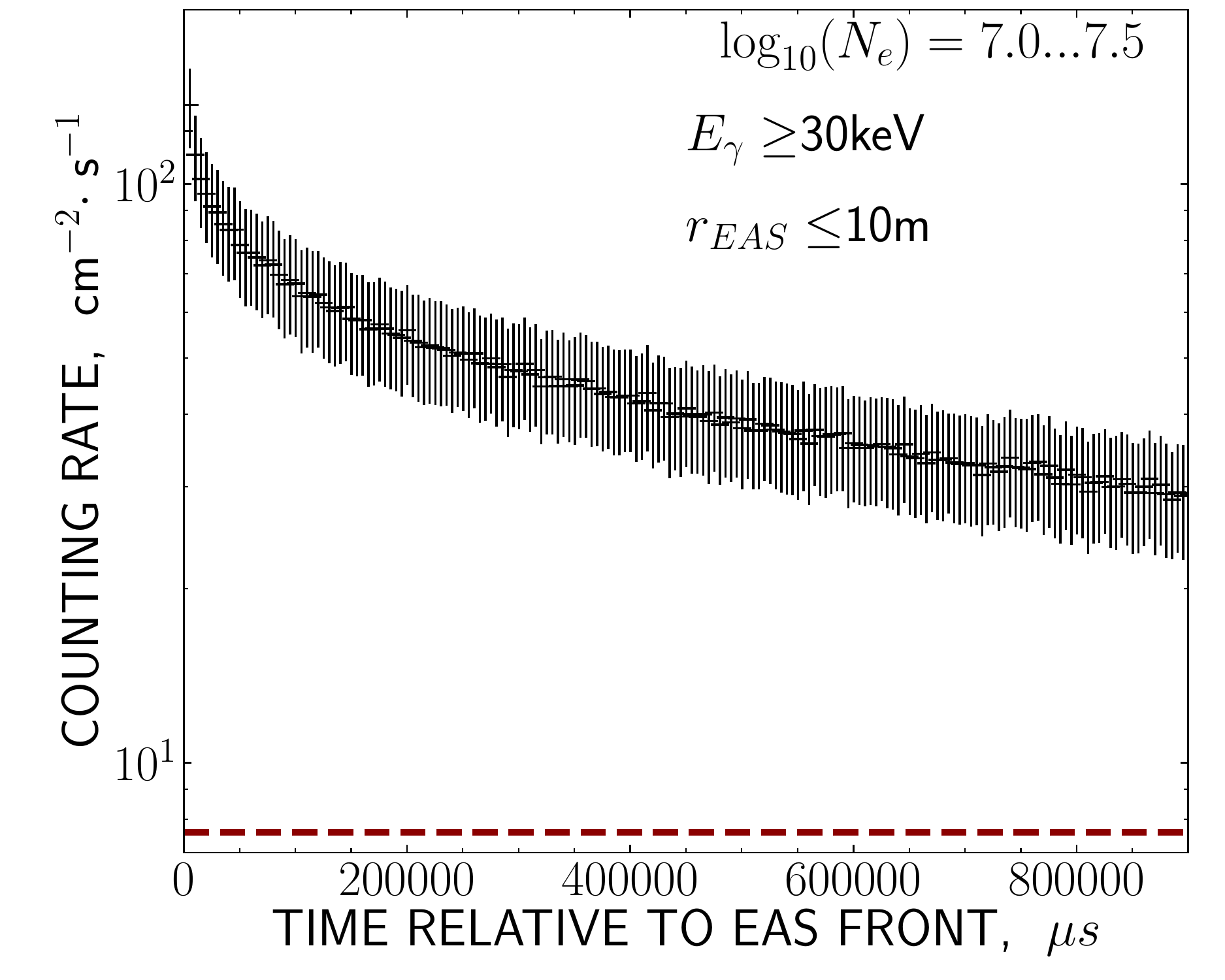}
\includegraphics[width=0.49\textwidth, trim=3mm 0mm 3mm 0mm]{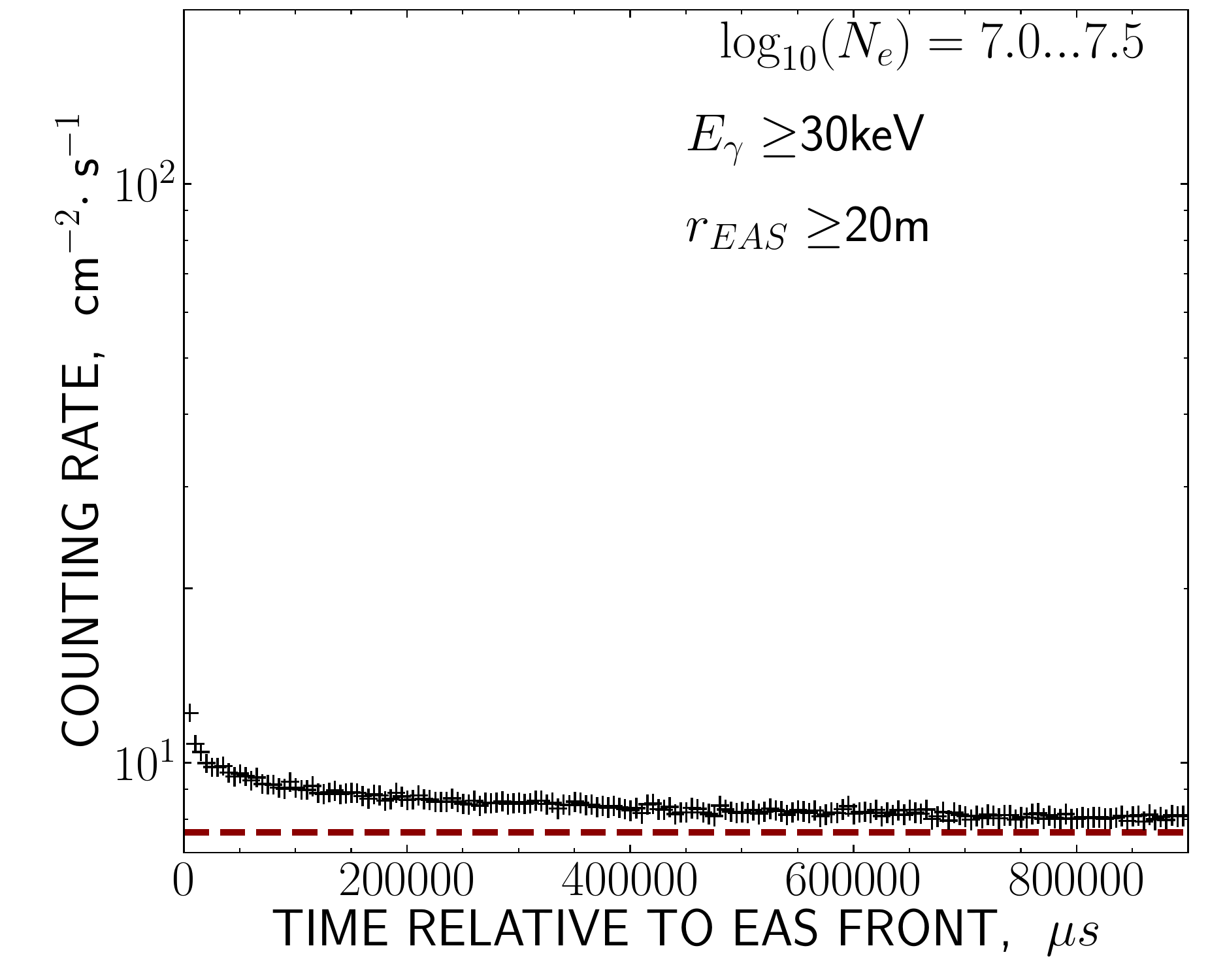}
\\
\includegraphics[width=0.49\textwidth, trim=3mm 0mm 3mm 0mm]{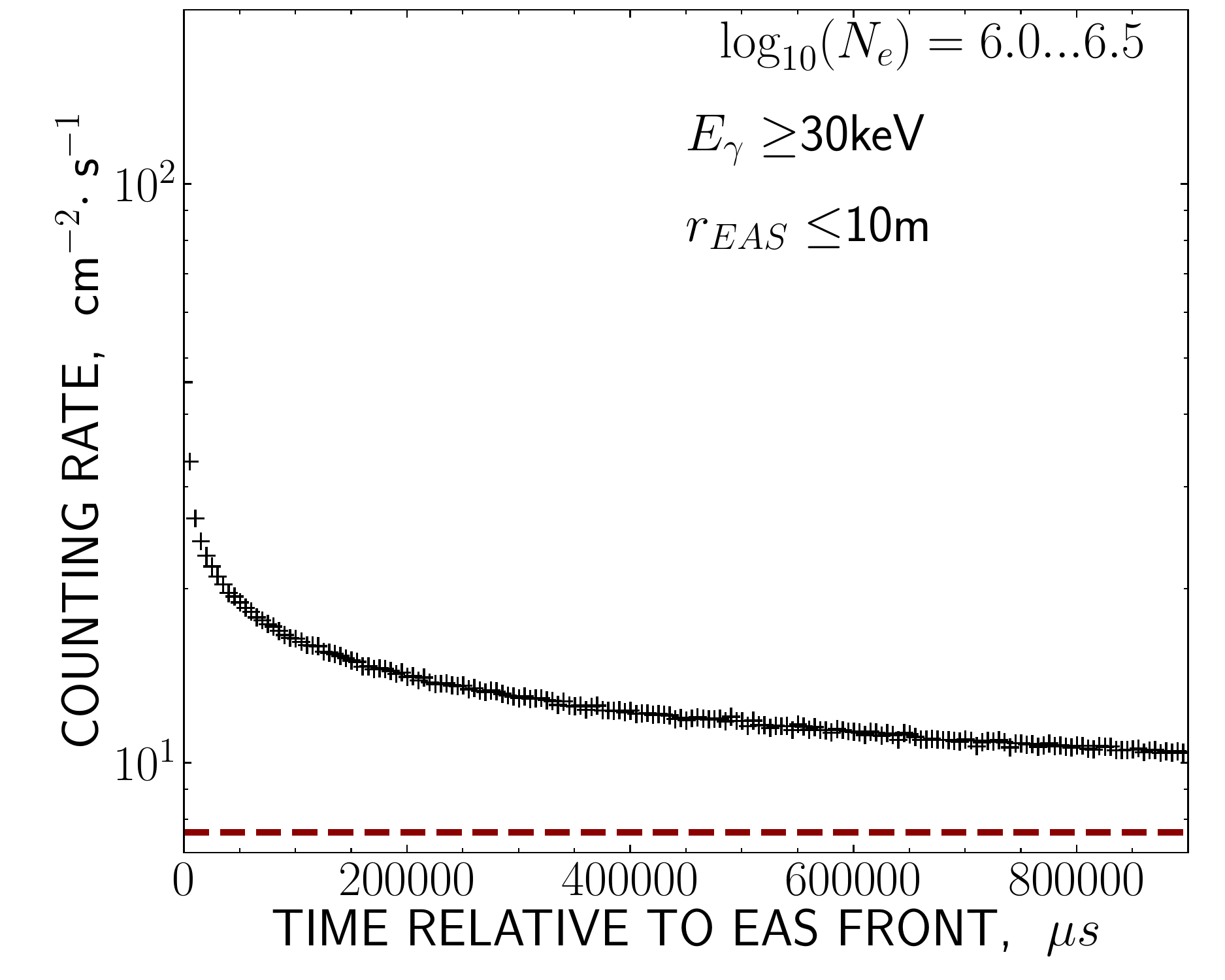}
\includegraphics[width=0.49\textwidth, trim=3mm 0mm 3mm 0mm]{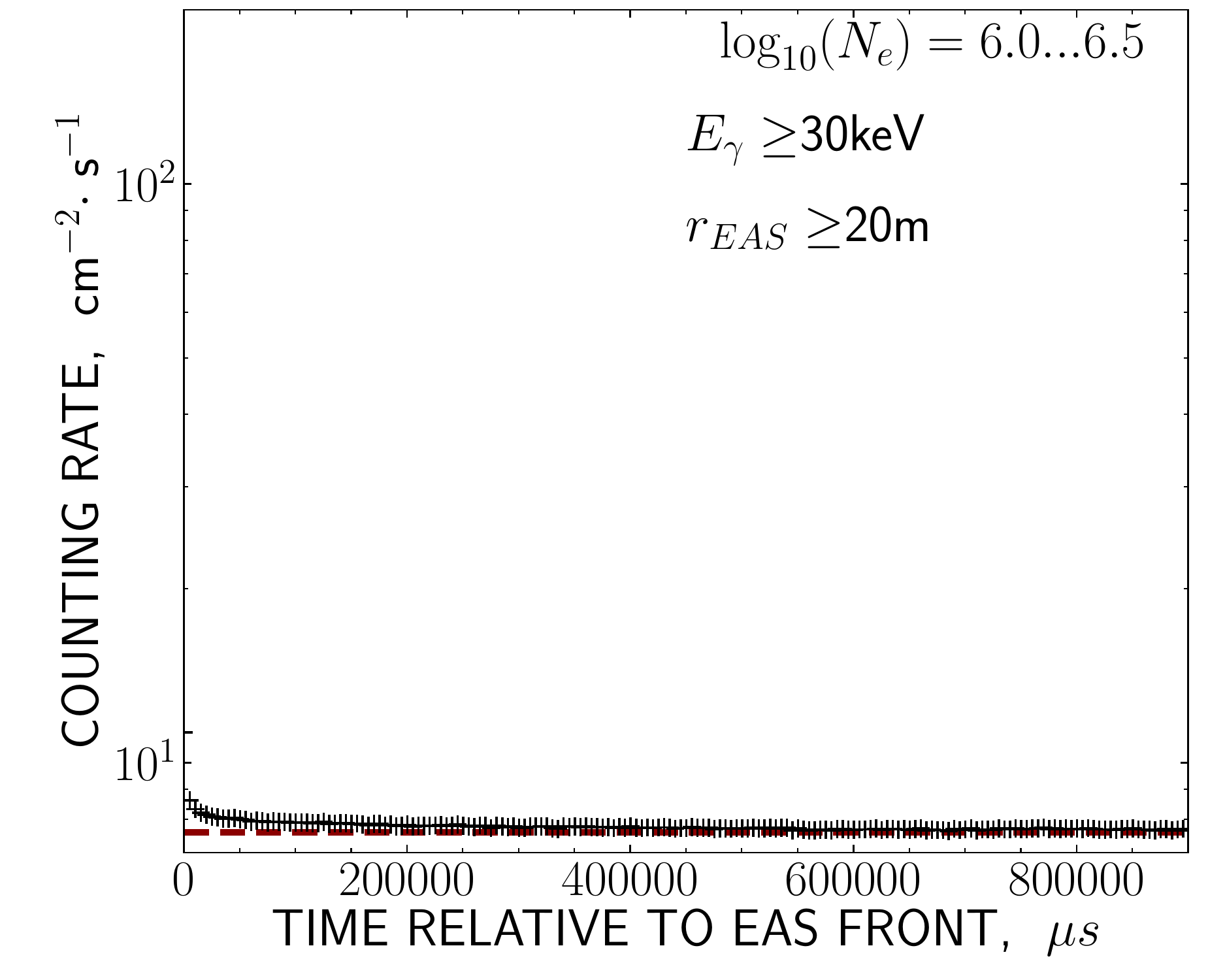}
%\\
\includegraphics[width=0.49\textwidth, trim=3mm 0mm 3mm 0mm]{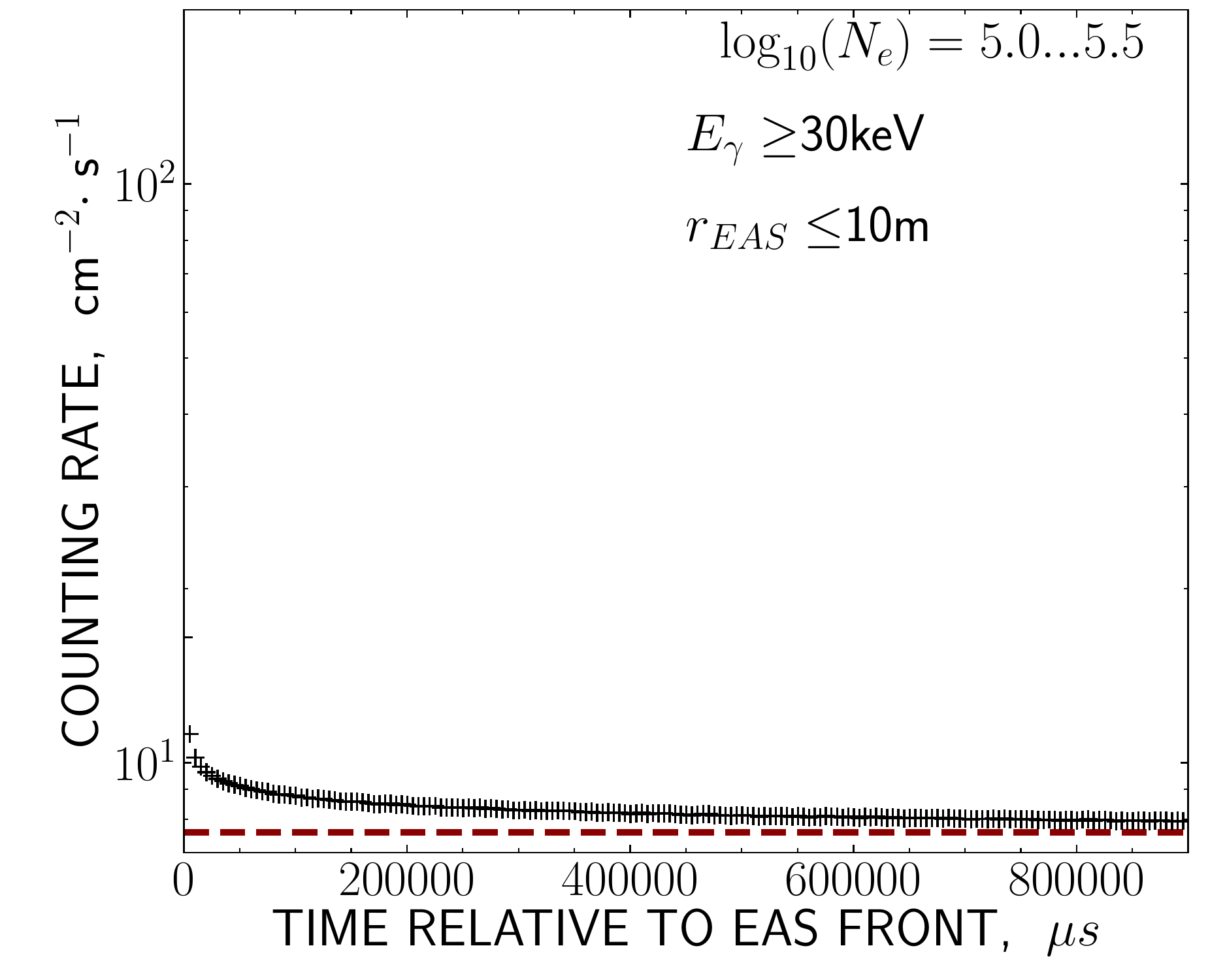}
\includegraphics[width=0.49\textwidth, trim=3mm 0mm 3mm 0mm]{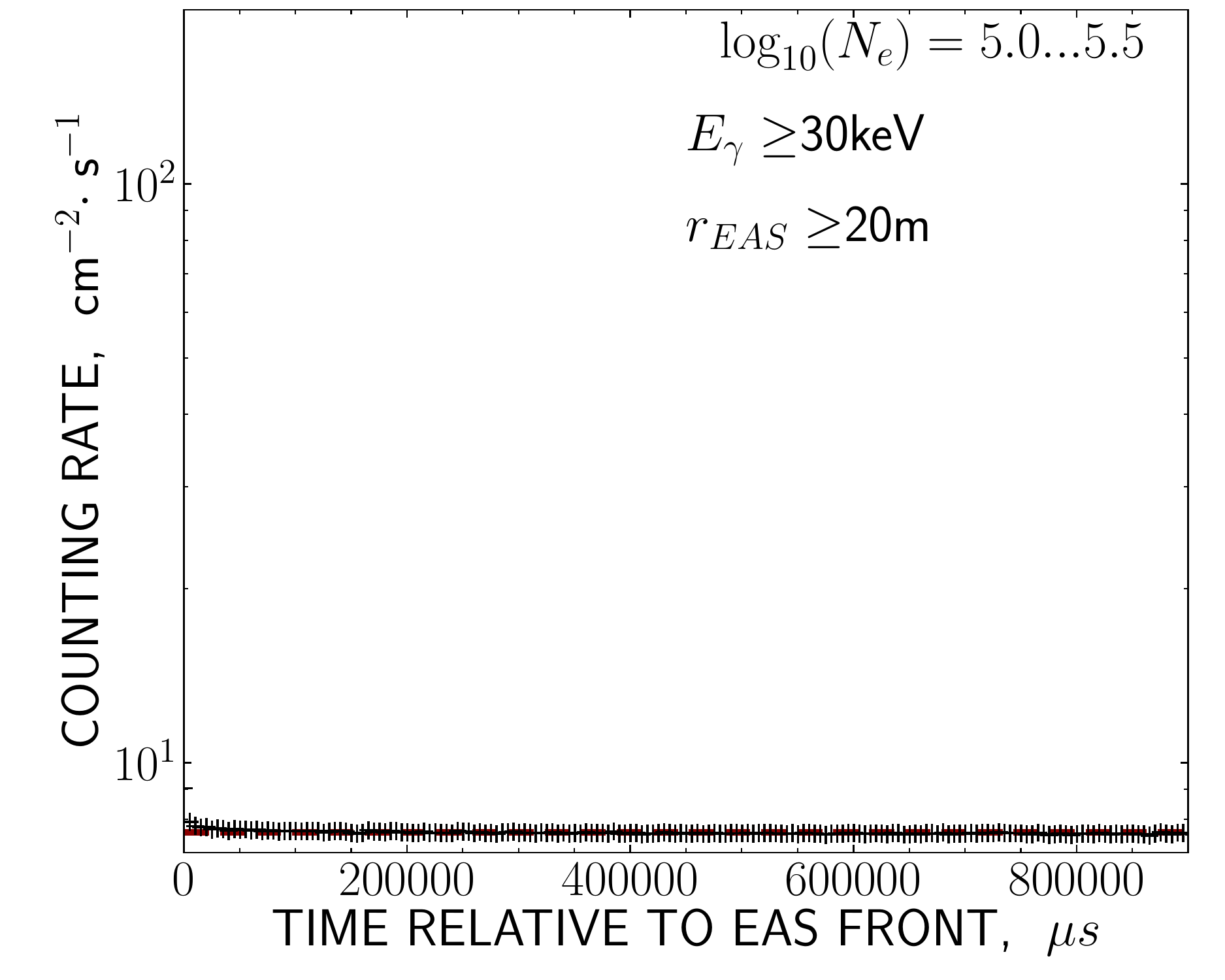}
\caption{Average time distributions of soft gamma radiation detected with the lowest energy threshold ($E_\gamma\geqslant 30$~keV) around the core ($r \leqslant 10$~m) and at periphery ($r \geqslant 20$~m) of the extensive air showers with different sizes $N_e$. The horizontal dashed lines mark the level of the background radiation intensity.}
\label{figigammotempo}
\end{center}
\end{figure*}

Figure~\ref{figigammotempo} demonstrates the mean temporal distributions of the intensity of soft radiation obtained by averaging of signal series which were registered in the first amplitude channel of gamma ray detector in various EAS events. Similarly to the case of neutron signals considered in Section~\ref{sectineutrotempo}, this averaging was done separately for some groups of EAS events with close values of their shower size parameter $N_e$ which have occurred at different distances $r$ from  gamma detector point: left column plots in Figure~\ref{figigammotempo} correspond to the central EAS region with $r\leqslant 10$\,m, while the right column ones---to the shower periphery  with $r\geqslant 20$\,m. All intensity curves shown in Figure~\ref{figigammotempo} were normalized to the total sensitive area of the detector's scintillation crystal (570\,cm$^2$), and to its efficiency (detection probability) in the considered range of  gamma ray energies. The latter values were defined by corresponding plot of Figure~\ref{figieffici} and listed in Table~\ref{tabgammothreshs}.

\begin{figure}
\begin{center}
\includegraphics[width=0.50\textwidth, trim=10mm 0mm 0mm 0mm]{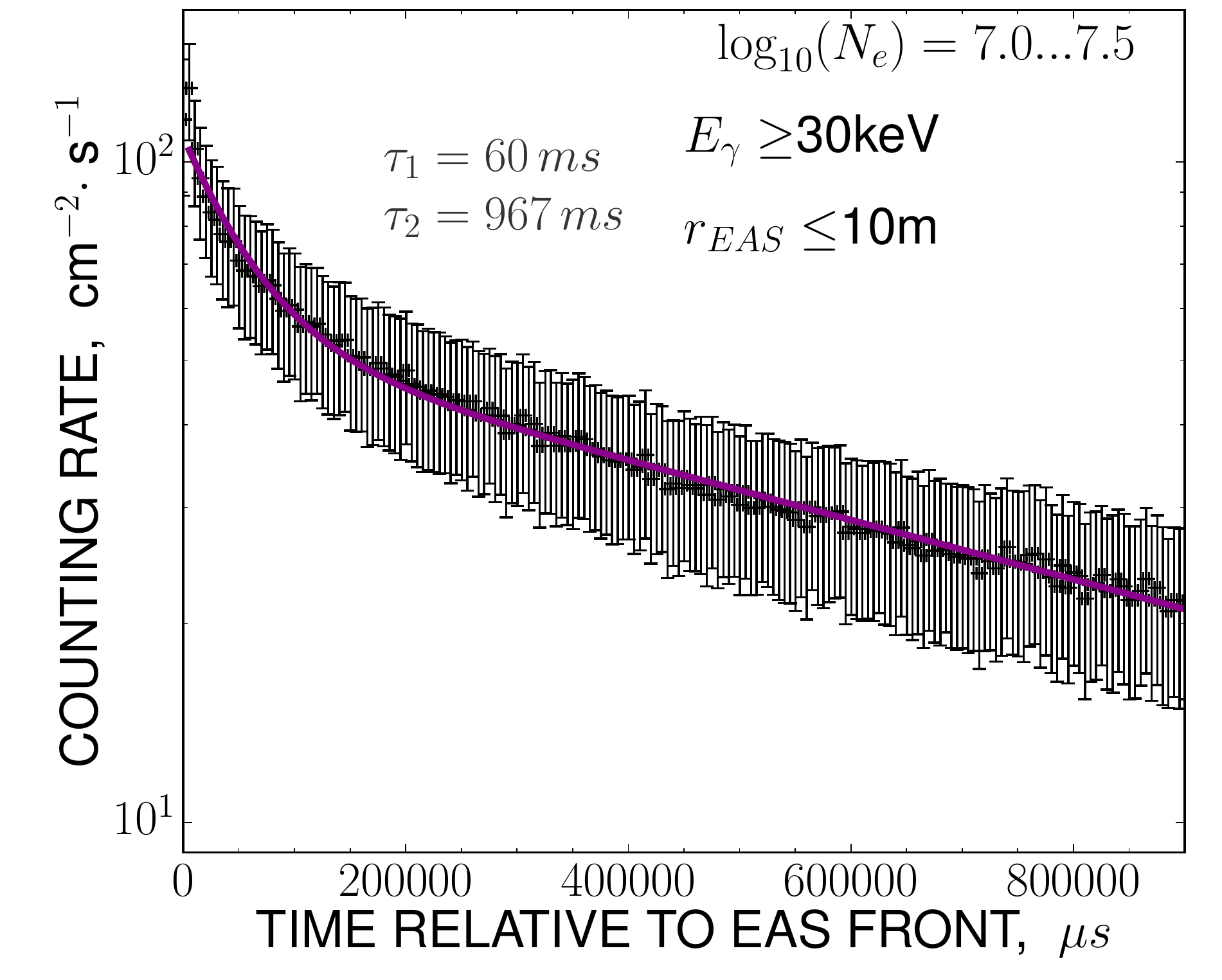}
\caption{Average time distribution of the intensity of soft radiation detected in central region of the large size EAS ($N_e\gtrsim 10^7$) and drawn after  subtraction of the background level, and with an exponential approximation of the type (\ref{equaneutrotau}) superimposed above the experimental points.}
\label{figigammotemponoback}
\end{center}
\end{figure}

The most interesting of the data presented in Figure~\ref{figigammotempo} is the intensity distribution for the large size EAS events ($N_e\gtrsim 10^7$) whose axes have come in vicinity to detector point: as it is obviously seen in the left top panel of this figure, even at expiration of a whole second long period the average intensity of the registered radiation still remains up to 3--4 times above its usual background. Pure deposit on the part of EAS connected gamma rays can be defined by subtraction of corresponding background level from the average distribution curve in the plot of Figure~\ref{figigammotempo}; as it is shown in the next Figure~\ref{figigammotemponoback}, the approximation of this deposit with a sum of two exponents of the type (\ref{equaneutrotau}) gives the lifetime value $\tau_2$ of about 1\,s.

Similar behaviour, though with a significantly less excess amplitude at large delay times, demonstrate in Figure~\ref{figigammotempo} the distributions build for the central part of much smaller EAS: those with $N_e\approx 10^6$ and $N_e\approx 10^5$.

\begin{table*}
\caption{\label{tabgammotempo} Average deposit into total flux of the registered $E_\gamma\geqslant 30$~keV radiation on the part of EAS connected gamma rays within the shower core region ($r\leqslant 10$~m), $\Delta {R(t)}= {R(t)}- {R}_{bckgr}$ (in the units of~cm$^{-2}$s$^{-1}$), in dependence on the delay time $t$ since shower passage.}
\begin{center}
\begin{tabular*}{\textwidth}{@{\extracolsep{ \fill } } cccccccc }
\hline
$\log_{10}(N_e)$&
t=10\,ms&                  %1
t=100\,ms &          %2
t=200\,ms&         %3
t=300\,ms&         %4
t=400\,ms&         %5
t=700\,ms&        %6
t=1000\,ms \\      %7
\hline

$7.0\ldots7.5$&
240&                   %1 10
60&                  %2 100
49&                  %3 200
42&                  %4 300
30&                   %5 500
26&                   %6 700
19\\                 %7 1000

$6.0\ldots6.5$&
21&                   %1 10
10&                   %2 100
6.6&                  %3 200
5.6&                   %4 300
4.5&                  %5 500
3.6&                  %6 700
2.9 \\                %7 1000

$5.0\ldots5.5$&
2.8&                   %1 10
1.4&                   %2 100
1.1&                   %3 200
0.9&                  %4 300
0.7&                  %5 500
0.6&                  %6 700
0.5 \\                %7 1000
\hline
\end{tabular*}
\end{center}
\end{table*}

\begin{figure*}
\begin{center}
\includegraphics[width=0.49\textwidth, trim=0mm 0mm 0mm 0mm]{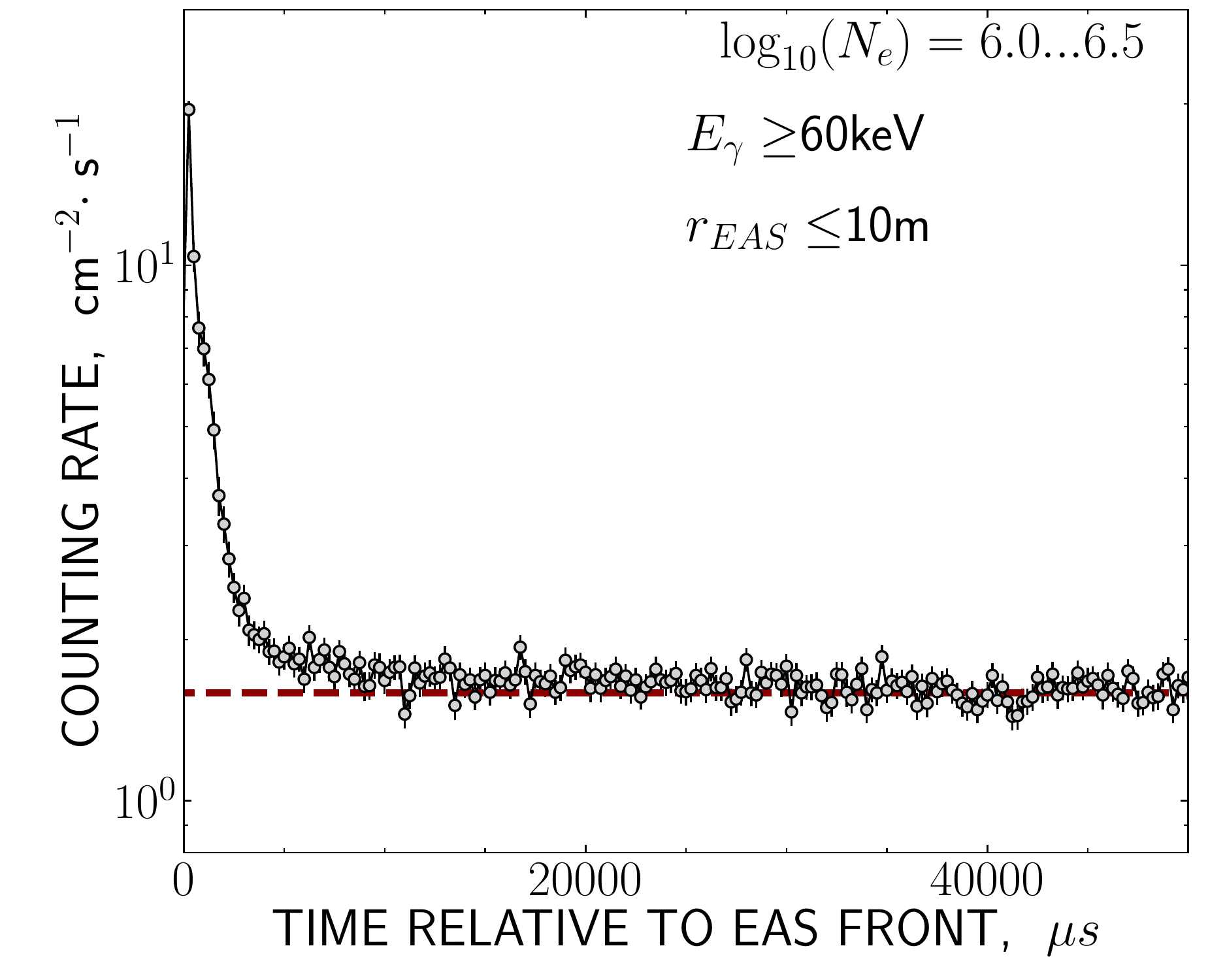}
%\\
\includegraphics[width=0.49\textwidth, trim=0mm 0mm 0mm 0mm]{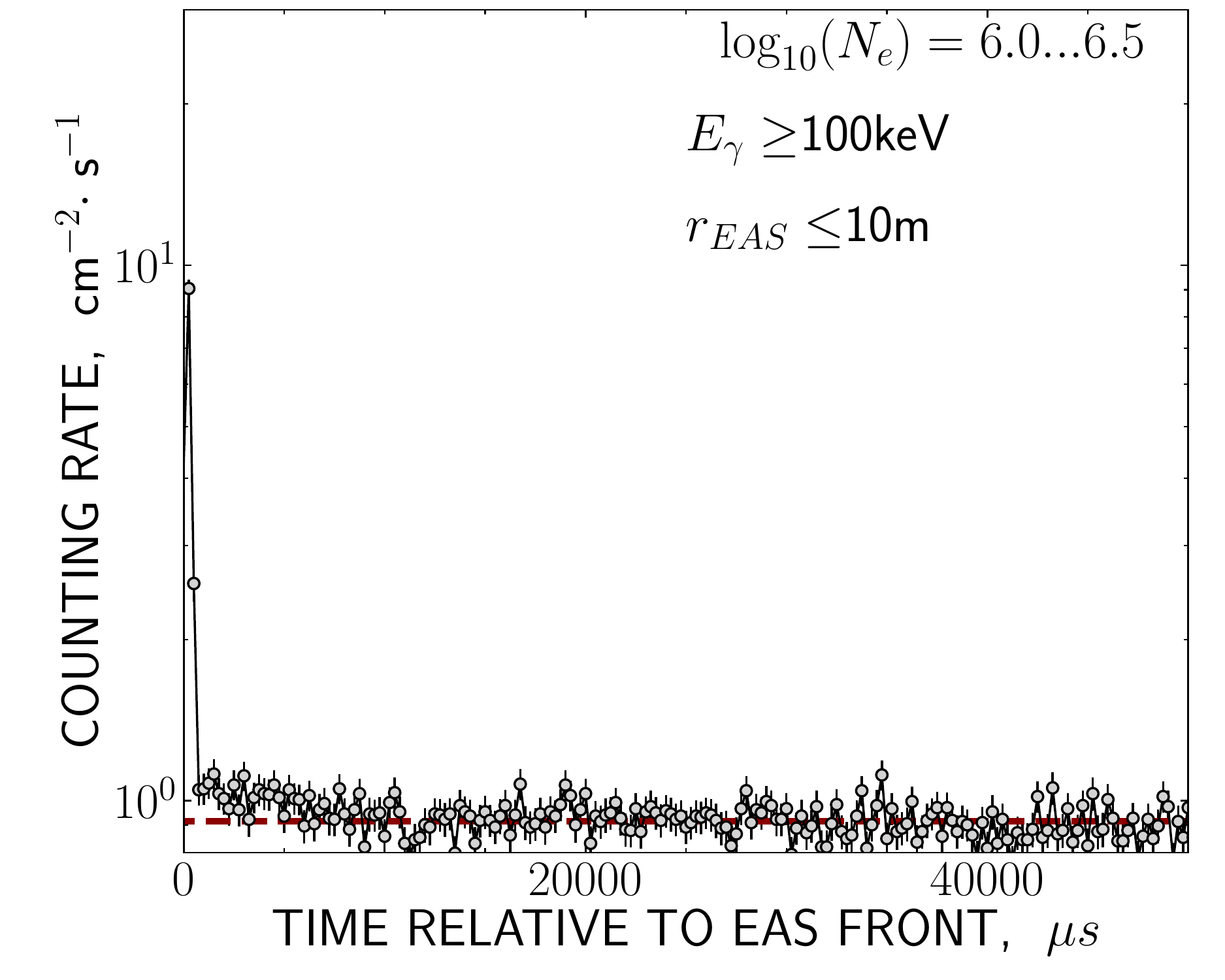}
\caption{Average distributions of the gamma ray intensity detected within the central EAS region ($r \leqslant 10$~m) with higher energy thresholds.}
\label{figigammotempohigherthresholds}
\end{center}
\end{figure*}

Absolute deposit of the EAS connected gamma rays into the total flux of the detected radiation can be calculated as a difference between the amplitude $R(t)$ of the average intensity curves in Figure~\ref{figigammotempo}, and the mean level of  background counting rate:
\begin{equation}
\Delta  {R(t)}= {R(t)}- {R}_{bckgr}.
\label{equadeltai}
\end{equation}
In case of the lowest energy threshold of the detected gam\-ma-radiation ($E_\gamma \geqslant 30$\,keV) the differences $\Delta {R(t)}$ are listed in Table~\ref{tabgammotempo}. It is seen in this table that within the central region of a large size EAS the flux of accompanying soft radiation varies from its peak value of about $200-250$\,cm$^{-2}$s$^{-1}$ which is typical shortly after the sho\-wer passage, down to $\Delta  {R} \approx 20$\,cm$^{-2}$s$^{-1}$ at the very end of the considered time period. With decreasing the size range of the selected EAS the intensity of delayed radiation falls rapidly down, but close to the shower axis it still remains quite detectable even in the showers with $N_e\approx 10^5$.

In contrast to this, from the right column plots of  Figure~\ref{figigammotempo} it follows that at the shower periphery some residual radiation over the late delay time can be found only in the large size events with $N_e\gtrsim 10^7$. After smaller showers with $N_e\approx 10^{5}-10^6$ any surplus flux above the usual background is absent just at a few tens---a hundred of milliseconds order times since the passage of shower front. Similarly to that, intensity distributions from Figure~\ref{figigammotempohigherthresholds} demonstrate a disappearance of any noticeably delayed signal even in central region of the showers with the rise of the minimum energy threshold of the detected radiation: just in the case of $E_\gamma \geqslant 60$\,keV condition the counting rate falls to its background level after a few tens of milliseconds, and with $E_\gamma \geqslant 100$\,keV threshold a single intensity outburst remains only at the passage of the main front of shower particles.

Thus, from the data presented here it follows that any nontrivial temporal behaviour of the EAS related radiation is connected with the soft gamma ray component of a few tens of keV order energy. A preferable spatial region for observation of such effects is confined within the closest neighbourhood of the shower center, and the probability to reveal any significantly delayed flux of soft gamma radiation after the EAS passage grows quickly with the size of the corresponding shower.

\subsection{Integral fluence of soft gamma radiation}

\begin{figure}
\begin{center}
\includegraphics[width=0.5\textwidth, trim=18mm 0mm 6mm 0mm]{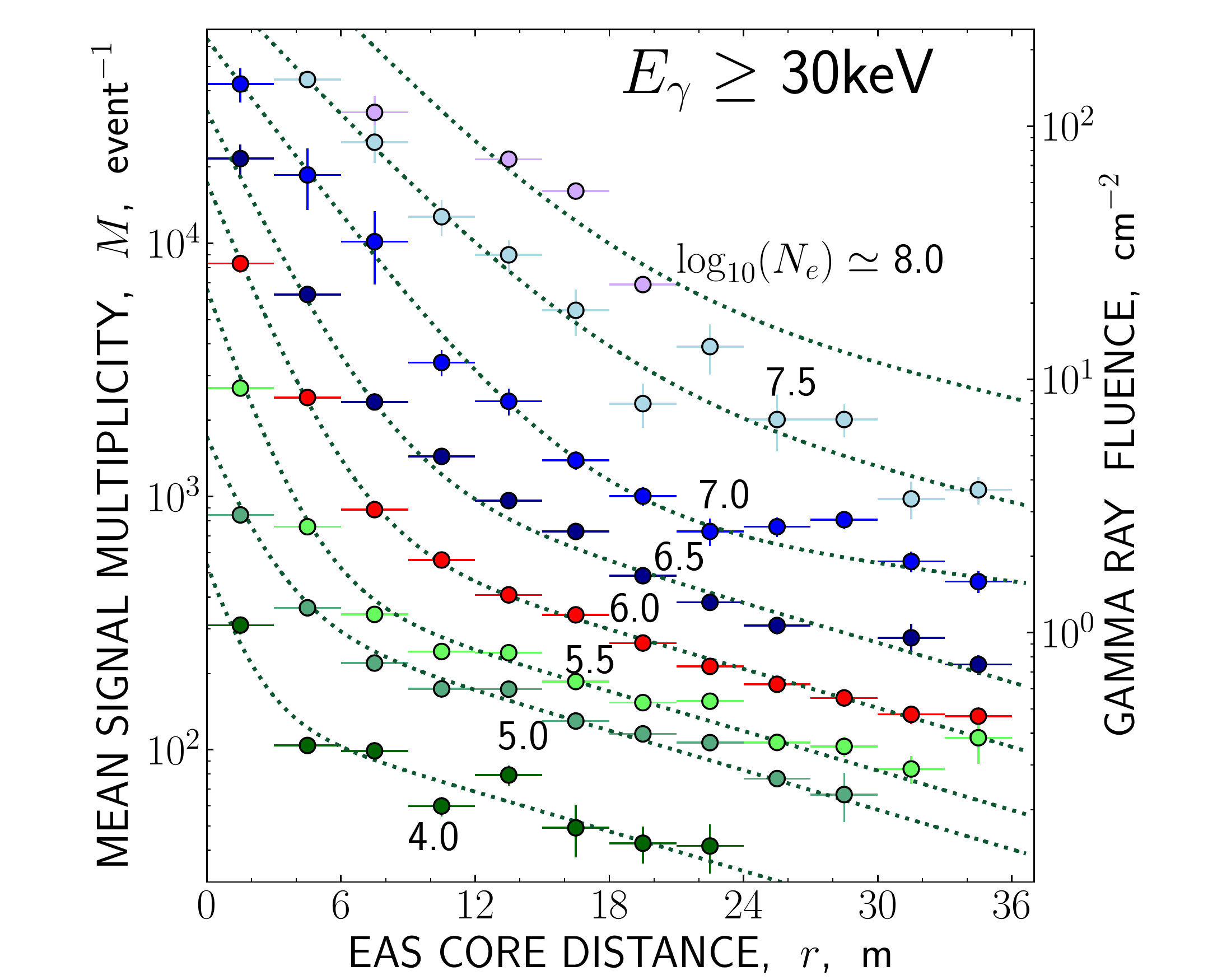}
\caption{
Spatial distributions of the average pulse signal multiplicity $M$ accepted in lowest threshold channel ($E_\gamma \geqslant 30$keV) of the gamma ray detector during the next second after the passage of extensive air shower. The numbers beside curves mean the decimal logarithm value of the average size $N_e$ of corresponding EAS; dotted lines mark an exponential approximation (\ref{equarho}) of the experimentally detected points; background is subtracted. Right axis is graduated in the units of the local integral fluence of gamma rays which was calculated for respective $M$ values with account to the sensitive area and detection efficiency of the gamma ray detector.
}
\label{figigammoldf}
\end{center}
\end{figure}

The total fluence of gamma rays born as a consequence of EAS passage can be calculated similarly to the case of neutron signals which was considered in Section~\ref{sectineutromulti}: taking into account the size $N_e$ and axis position parameters for each EAS, to split the whole statistics of the detected shower events into a number of groups with close values of their size parameter $N_e$ and distance $r$ between the shower center and gamma detector, and to calculate subsequently an average multiplicity $M$ of detector pulse signals with a given energy threshold $E_\gamma$ of detected gamma rays for particular combinations of $(r,N_e)$. Then, a family of the mean spatial distribution functions $M(r)$ can be drawn for various ranges of $N_e$ value. In Figure~\ref{figigammoldf} the results of this procedure are presented for the most interesting case of the gamma ray detector channel with the lowest energy threshold ($E_\gamma \geqslant 30$keV).

The gate time $T_g$ to calculate the multiplicity values presented in Figure~\ref{figigammoldf} was equal to the whole one second long period which was initially accepted in the considered experiment for signal selection after EAS passage. With $\tau_1\approx 60$\,ms and $\tau_2\approx 970$\,ms lifetime estimations which result for two components of an exponential approximation $I(t)$ in Figure~\ref{figigammotemponoback}, such a duration of the gate time ensures collection of about \sfrac{2}{3}~fraction part of the EAS connected gamma ray signals:
\begin{equation}
\int_{0}^{ {T_g}}I(t)dt \biggm/ \int_0^\infty I(t)dt \approx 0.67.
\label{equagammofrati}
\end{equation}
% see the 'limited_gate_time_error_script' label!
% integral for the limits    0...\infty: 54955145.68016022,
% integral for the limits    0...1.e^6: 36619488.143313706
% relation: 36619488/54955145 = 0.6663523133275329
The background deposit to be subtracted from the ``raw'' pulse numbers which have been registered over the whole ${T_g}$ gate time can be defined as a product ${T_g}\cdot{R}_{bckgr}$, with the background counting rates ${R}_{bckgr}$ taken from  Table~\ref{tabgammothreshs}. Actually, the points in Figure~\ref{figigammoldf} represent the multiplicity values $M$ which remain after such a background subtraction, and thus relate exclusively to the gamma ray signal connected with EAS passage. In this case the total radiation fluence at a distance $r$ from the shower axis is $F_\gamma(r) = M(r)/(S\cdot \varepsilon)$, where $S$ is the surface area of scintillation crystal, and $\varepsilon$ is the detector efficiency in the considered range of radiation energy. For convenience, the right axis of Figure~\ref{figigammoldf} is graduated immediately in the units of $F_\gamma$ (with efficiency value $\varepsilon=0.5$ which follows from Table~\ref{tabgammothreshs} for an energy threshold range of $E_\gamma\geqslant 30$\,keV).

\begin{figure}
\begin{center}
\includegraphics[width=0.50\textwidth, trim=0mm 0mm 15mm 0mm]{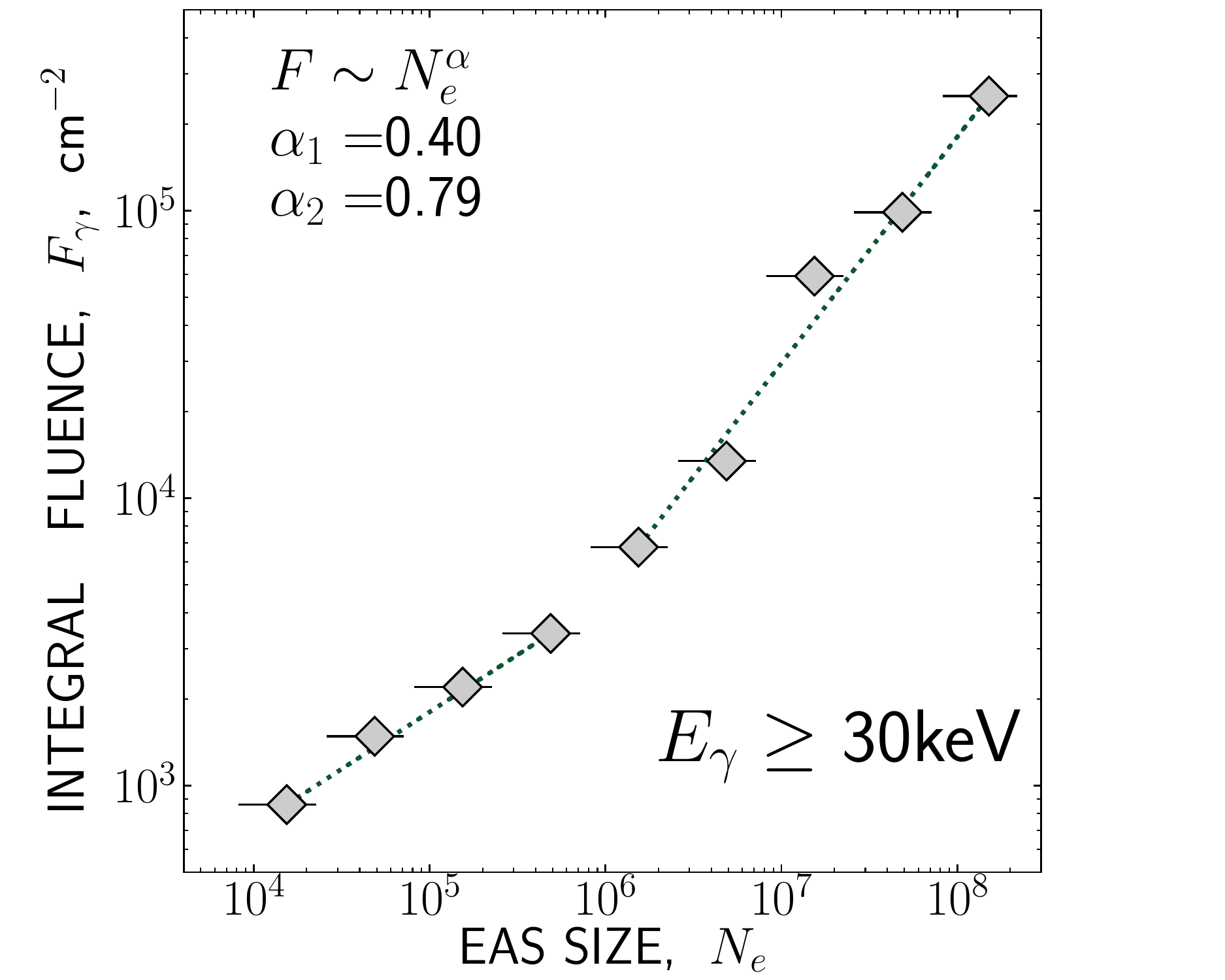}
\caption{
Mean total fluence of the soft gamma radiation $F_\gamma(\geqslant$30~keV$)$ detected after EAS passages in the whole second long gate time, and presented in dependence on the average size $N_e$ of the corresponding showers. Dotted lines mark a piecewise power approximation of the experimental points.
}
\label{figigammomulti}
\end{center}
\end{figure}

As it is seen in Figure~\ref{figigammoldf}, generally  spatial distributions of the average multiplicity number $M(r)$ follow a sum of exponents like function (\ref{equarho}) which was introduced above in Section~\ref{sectineutromulti}. A number of approximating functions $D_\gamma(r)$  of such exponential type which have been obtained by fitting experimental points $M(r)$ in various $N_e$ range distributions are presented in Figure~\ref{figigammoldf} with dotted lines. Since the largest value of spatial scale parameter $\rho$ in the formula (\ref{equarho}) occurs being merely of about $15-20$\,m for all these approximations, a comparatively narrow range of the distance argument $r$ variation along the abscissa axis in Figure~\ref{figigammoldf} happens to be quite sufficient to trace adequately the spatial distribution of the EAS connected gamma rays.
(In turn, the limitation of the distance range in Figure~\ref{figigammoldf} is a consequence of the restriction of all the considered statistics only with those EAS whose axes were hitting the area of the central ``carpet'' of the detector system for a better definition of shower parameters).

Numerical integration of the approximation functions $D_\gamma(r)$ between zero and infinity limits, like the equation~(\ref{equaoitegrho}) above, permits to calculate the average total fluence $F_\gamma$ of the soft gamma radiation which has come through the area of the detector system during the whole second long gate time after the passage of EAS. The resulting fluence values $F_\gamma$ are plotted in Figure~\ref{figigammomulti} depending on the average size value $N_e$ of the showers. As it follows from the equation (\ref{equagammofrati}), the accepted duration of the gate time leaves uncontrolled a rather noticeable 30\% portion of all gamma ray signals at the tail of their temporal distribution, so the fluence values designated in Figure~\ref{figigammomulti} most likely should be considered as a possible lower limit of corresponding estimations, especially in the region of large size EAS.

As it has been done earlier in  Section~\ref{sectineutromulti} for the case of neutron signal, the gamma ray fluence points in Figure~\ref{figigammomulti} can be fitted with a piecewise power function of $F_\gamma\sim N_e^\alpha$ type, and such approximation is represented in this figure with a pair of dotted lines. As it is seen, the power index $\alpha$ of this dependence experiences a sharp, twice as much change of its value around the point $N_e\approx 10^{6}$, that is just at the same EAS size border as the neutron flux does in the similar plot of  Figure~\ref{figineutromulti}. As it was discussed in Section~\ref{sectineutromultiknee}, at the height of the Tien Shan mountain station it is this range of EAS size values which corresponds to the position of the $3\cdot 10^{15}$\,eV knee in the primary spectrum of cosmic ray particles. Hence, the observed change in the behaviour of $F_\gamma(N_e)$ dependence of EAS accompanying soft gamma radiation can probably be considered as one more non-trivial effect found in the same energy range of the primary cosmic rays spectrum.

\section{Conclusion}

All presented experimental results on the low-energy neutron and gam\-ma-ray accompaniment of the $(10^{14}-10^{17})$~eV extensive air showers can be summarized as follows: %the following:
\begin{itemize}
\item
within the central EAS region (at a $5-10$\,m order distance from the shower axis) the fluence of EAS connected neutrons varies in the limits of $(10^{-4}-1)$~cm$^{-2}$ depending on the size of the shower, as it follows from Figure~\ref{figineutrofpr}, and mostly all these neutrons belong to the thermal range of kinetic energy;

\item
according to Figure~\ref{figineutromulti}, the average integral fluence of EAS connected neutron flux has a piecewise power dependence on the shower size parameter $N_e$, with a sharp, twice as much increase of its power index around the value of $N_e\approx 10^6$;

\item
it was revealed an extremely prolonged flux of soft gamma radiation which exists up to a few seconds order times after the passage of shower front; this effect is mainly confined within the spatial region of the EAS core where the local fluence of delayed gamma rays $F_\gamma$ can reach the values of up to $\sim$$10^3$\,cm$^{-2}$ in the case of large size showers ($N_e\approx 10^8$), and most prominently it reveals itself among the soft ($30-50$)\,keV gamma radiation component (see Figures~\ref{figigammotempo}--\ref{figigammotempohigherthresholds});

\item
after the integration of the average spatial distributions  $F_\gamma(r)$ over the whole area of EAS cross-section, the resulting integral fluence of accompanying soft radiation in Figure~\ref{figigammomulti} follows, again, a piecewise power dependence on the average shower size $N_e$, with doubling of its power index around the value of $N_e\approx 10^6$.
\end{itemize}

As to the origin of delayed EAS accompaniment, the detected neutron flux should evidently arise in the interaction of nuclear active EAS particles with the matter of the outer environment, which thus plays a role of a peculiar hadron ca\-lo\-ri\-me\-ter. This effect can be of use for further study of interaction properties of the high energy hadronic components of natural cosmic rays. Due to a comparatively wide scattering of the thermalized evaporation neutrons in the process of their diffusion such a technique can have some advantages since a considerable spatial areas of the order of $(10^2-10^3)$\,m$^{2}$ can be easily controlled by a rather limited set of neutron detectors.
%, primarily within the central region of extensive air showers, since the most energetic hadrons do usually concentrate around an EAS core.

Delayed flux of gamma radiation detected in this experiment after the passage of EAS front could possibly have a twofold origin. First, according to Figure~\ref{figigammotemponoback} there must be a relatively ``fast'' component with a lifetime estimation of about 60\,ms which is close to the  existence time of evaporation neutrons generated immediately at the EAS moment. Hence, these gamma rays could probably be ascribed to the radiative neutron captures in the surrounding environment. Another exponential component with a whole second order lifetime is likely to originate from radioactive gamma decays of the short-living exited nuclei which arise through activation of the environmental matter by EAS connected neutrons. Though both these suppositions should be a subject of further investigation, they de\-mon\-stra\-te once more the possibility to use the delayed radiation signals detected after the EAS passage as an effective probe to investigate the properties of the high energy nuclear processes by multiplicity of the thermalized evaporation neutrons and other low-energy radiation components produced in interaction of the EAS hadrons.

Irregularities newly found in dependence of the delayed neutron and gamma ray signal multiplicity on the average size parameter of the corresponding EAS should be considered especially as an example of such type of a study. As it was mentioned above, the shower size border value of $N_e\approx 10^6$ at which a sharp change resides in the power index of both components corresponds to the position of the $3\cdot 10^{15}$\,eV knee in the primary cosmic rays spectrum. On the other hand, any phenomenological effects observed experimentally in behaviour of the delayed neutron and gamma ray signals, as it was discussed in Section~\ref{sectineutromultiknee}, must reflect some basic peculiarity of the neutron production process which should arise within this energy range in interaction of EAS hadrons with the matter. The indication to threshold appearance in the $N_e\approx 10^6$ EAS of some additional energy transmission mechanism into a multitude of (possibly, low-energy) hadrons tightly concentrated around the shower axis can be of interest from the viewpoint of the knee explanation problem. Correspondingly, further development of the methods based on detection of delayed radiations with reduced energy threshold, and finally aimed at the study of specific patterns of hadron interaction in the EAS is anticipated in the plans of modern experimental activity at Tien Shan Mountain Cosmic Ray Station.

\section*{Acknowledgements}
This work was supported by scientific research programs \#BR05236291 and \#BR05236494 of the Ministry of Education and Science of Kazakhstan Republic.

%
% BibTeX users please use
% \bibliographystyle{}
% \bibliography{}
%
% Non-BibTeX users please use

\newcommand {\etal}{et al. }
% what's below was generated by bibtex.
\providecommand{\url}[1]{{#1}}
\providecommand{\urlprefix}{URL }
\expandafter\ifx\csname urlstyle\endcsname\relax
  \providecommand{\doi}[1]{DOI \discretionary{}{}{}#1}\else
  \providecommand{\doi}{DOI \discretionary{}{}{}\begingroup
  \urlstyle{rm}\Url}\fi

\end{document}